%% file: main.tex
\documentclass[aps,prl,reprint,amssymb,showpacs,onecolumn,superscriptaddress,notitlepage]{revtex4-2}
\usepackage{bm}
\usepackage{graphicx}
\usepackage{dcolumn}
\usepackage{braket}
\usepackage{natbib}
\usepackage{amsmath}
\usepackage{color}
\usepackage{ulem}
\usepackage{subcaption} 
\usepackage{siunitx}

\definecolor{gray}{rgb}{0.7,0.7,0.7}
\definecolor{orange}{rgb}{1, 0.4, 0}
\definecolor{dgreen}{rgb}{0.0, 0.4, 0.0}
\definecolor{yblue}{rgb}{0.06, 0.3, 0.57}

\newcommand\ldsout{\bgroup\markoverwith{\textcolor{blue}{\rule[0.5ex]{2pt}{0.4pt}}}\ULon}

\usepackage{subfiles}
\begin{document}
\title{Fabrication of Nanostructured GaAs/AlGaAs Waveguide for Low-Density Polariton Condensation from a Bound State in the Continuum}
\author{F. Riminucci}
\affiliation{Molecular Foundry, Lawrence Berkeley National Laboratory, One Cyclotron Road, Berkeley, California 94720, USA}
\affiliation{Dipartimento di Matematica e Fisica, ‘Ennio de Giorgi’, Università del Salento, Lecce, Italy}
\author{V. Ardizzone}
\affiliation{CNR Nanotec, Institute of Nanotechnology, via Monteroni, 73100, Lecce}
\author{L. Francaviglia}
\affiliation{Molecular Foundry, Lawrence Berkeley National Laboratory, One Cyclotron Road, Berkeley, California 94720, USA}
\author{M. Lorenzon}
\affiliation{Molecular Foundry, Lawrence Berkeley National Laboratory, One Cyclotron Road, Berkeley, California 94720, USA}
\author{C. Stavrakas}
\affiliation{Molecular Foundry, Lawrence Berkeley National Laboratory, One Cyclotron Road, Berkeley, California 94720, USA}
\author{S. Dhuey}
\affiliation{Molecular Foundry, Lawrence Berkeley National Laboratory, One Cyclotron Road, Berkeley, California 94720, USA}
\author{A. Schwartzberg}
\affiliation{Molecular Foundry, Lawrence Berkeley National Laboratory, One Cyclotron Road, Berkeley, California 94720, USA}
\author{S. Zanotti}
\affiliation{Dipartimento di Fisica, Universit\`a di Pavia, via Bassi 6, Pavia (IT)}
\author{D. Gerace}
\affiliation{Dipartimento di Fisica, Universit\`a di Pavia, via Bassi 6, Pavia (IT)}
\author{K. Baldwin}
\affiliation{PRISM, Princeton Institute for the Science and Technology of Materials, Princeton University, Princeton, New Jersey 08540, USA}
\author{L. N. Pfeiffer}
\affiliation{PRISM, Princeton Institute for the Science and Technology of Materials, Princeton University, Princeton, New Jersey 08540, USA}
\author{G. Gigli}
\affiliation{Dipartimento di Matematica e Fisica, ‘Ennio de Giorgi’, Università del Salento, Lecce, Italy}
\affiliation{CNR Nanotec, Institute of Nanotechnology, via Monteroni, 73100, Lecce}
\author{D. F. Ogletree}
\affiliation{Molecular Foundry, Lawrence Berkeley National Laboratory, One Cyclotron Road, Berkeley, California 94720, USA}
\author{A. Weber-Bargioni} 
\affiliation{Molecular Foundry, Lawrence Berkeley National Laboratory, One Cyclotron Road, Berkeley, California 94720, USA}
\author{S. Cabrini}
\affiliation{Molecular Foundry, Lawrence Berkeley National Laboratory, One Cyclotron Road, Berkeley, California 94720, USA}
\author{D. Sanvitto}
\email{daniele.sanvitto@nanotec.cnr.it}
\affiliation{CNR Nanotec, Institute of Nanotechnology, via Monteroni, 73100, Lecce}

\begin{abstract}
Exciton-polaritons are hybrid light-matter states that arise from strong coupling between an exciton resonance and a photonic cavity mode. As bosonic excitations, they can undergo a phase transition to a condensed state that can emit coherent light without a population inversion. This aspect makes them good candidates for thresholdless lasers, yet short exciton-polariton lifetime has made it difficult to achieve condensation at very low power densities. In this sense, long-lived symmetry-protected states are excellent candidates to overcome the limitations that arise from the finite mirror reflectivity of monolithic microcavities. In this work we use a photonic symmetry protected bound state in the continuum coupled to an excitonic resonance to achieve state-of-the-art polariton condensation threshold in GaAs/AlGaAs waveguide. Most important, we show the influence of fabrication control and how surface passivation via atomic layer deposition provides a way to reduce exciton quenching at the grating sidewalls.
\end{abstract}

\maketitle
\section{Introduction}
Efficient low-threshold lasing is a long-sought goal for various applications spanning from integrated circuits to biological sensing \cite{Ma2019ApplicationsNanolasers,Hill2014AdvancesLasers}. However, as of today the emission of coherent and monochromatic light has relied on population inversion, which requires to overcome a threshold excitation power to achieve lasing. A different and new approach exploits hybrid light-matter particles, known as exciton-polaritons\cite{Weisbuch1992,article}, which arise in a semiconductor when the energy exchange rate between a photon and an exciton is higher than their losses. These particles possess properties inherited from their photonic component, such as a small effective mass, and from their excitonic component such as high nonlinearities\cite{Walker2019,Estrecho2019} that can be further enhanced via dipolar interactions\cite{Togan2018EnhancedPolaritons,Suarez-Forero2021EnhancementInteractions}. All these properties make them interesting candidates for devices\cite{Sanvitto2016} such as optical transistors\cite{Ballarini2013b,Sturm2014All-opticalInterferometer}, fast optical switches\cite{Zasedatelev2021Single-photonTemperature,Suarez-Forero2021UltrafastWaveguides}, and electrically-injected light sources \cite{Schneider2013}.\\ 
Because of their bosonic nature, they can undergo a phase transition to a coherent state, known as Bose-Einstein condensate\cite{Kasprzak2006,Byrnes2014} (BEC), without the need for population inversion, making thresholdless lasing theoretically possible\cite{Imamoglu1996}. 
A necessary condition for this phase transition to happen is that the rate of scattering processes\cite{Miesner1998BosonicCondensate} that populate the BEC state must exceed the loss rates\cite{Sanvitto2016}. This has been one of the main limitations in the historical achievement of exciton-polariton condensates, as polaritons tend to accumulate in the so-called  "bottleneck"\cite{Tassone1997BottleneckPolaritons,Richard2005ExperimentalPolaritons}. Such limitation was overcome\cite{Kasprzak2006} by growing highly reflective mirrors with layers exceeding 40 pairs in both sides of a microcavity as well as using many stacks of QWs.
In the last few years a new approach has enabled the achievement of polariton states in horizontal platforms\cite{Jamadi2018,Jamadi2019CompetitionMicrocavities,Walker2013,Walker2019,DiPaola2021Ultrafast-nonlinearTemperature,Suarez-Forero2020}, making the fabrication easier and less time consuming. A similar waveguide configuration with multiple QWs was used to investigate the enhancement of dipolar polariton interactions\cite{Rosenberg2016,Rosenberg2018,Liran2018,Suarez-Forero2021EnhancementInteractions}. More recently, the same heterostructure led to the achievement of a Bose-Einstein condensate from a BIC\cite{Ardizzone2021PolaritonContinuum} while also proving the topological charge possessed by the condensate. \\
In this work we show the achievement of a low-density condensate in a GaAs/AlGaAs waveguide by tuning the parameters of a low-loss shallow grating and then we study the effect of the reduction of excitonic losses on the threshold. 
In order to reduce the radiative losses we make use of a topologically protected bound state in the continuum \cite{Kodigala2017LasingContinuum,Hsu2016BoundContinuum,Doeleman2018ExperimentalContinuum} (BIC) which is introduced in the polariton dispersion through the realization of a 1D etched grating\cite{Kravtsov2020}. Since the BIC has an antisymmetric profile, it cannot couple to outgoing plane waves, resulting in a perfectly dark state with zero radiative linewidth \cite{Zhen2014TopologicalContinuum}. The exciton-polariton dispersion can be modified by changing the grating periodicity, which defines the energy at which the photonic modes cross. Such crossing can then be adjusted to bring the BIC closer to or further from the exciton, changing its excitonic fraction. This will ultimately change the threshold since the excitonic fraction has also an effect on the polariton relaxation process towards the BIC. This control is something that has no counterparts in classical vertical microcavities and can be tuned with extreme precision on several different gratings on the same chip. The excitonic fractions were estimated throughout this work by using the coupled oscillators model described in the supplementary materials.\\ A horizontal configuration and such long-lived state can drastically ease the fabrication with respect to vertical microcavities in order to achieve thermalization of polaritons into a single quantum state.
Here we demonstrate the achievement of low-threshold condensate thanks to simple, yet extremely effective sample processing. We will study three different shallow etched 1D gratings: 40nm, 90nm, and 130nm deep grooves. Finally, we show the reduction of the exciton-polariton linewidth and condensate threshold by passivating the trap states that form on the groove sidewalls during the etching process.

\section{Results}
\begin{figure*}[t!]
    \centering
    \begin{subfigure}[t]{0.51\textwidth}
        \centering
        \includegraphics[scale=0.24]{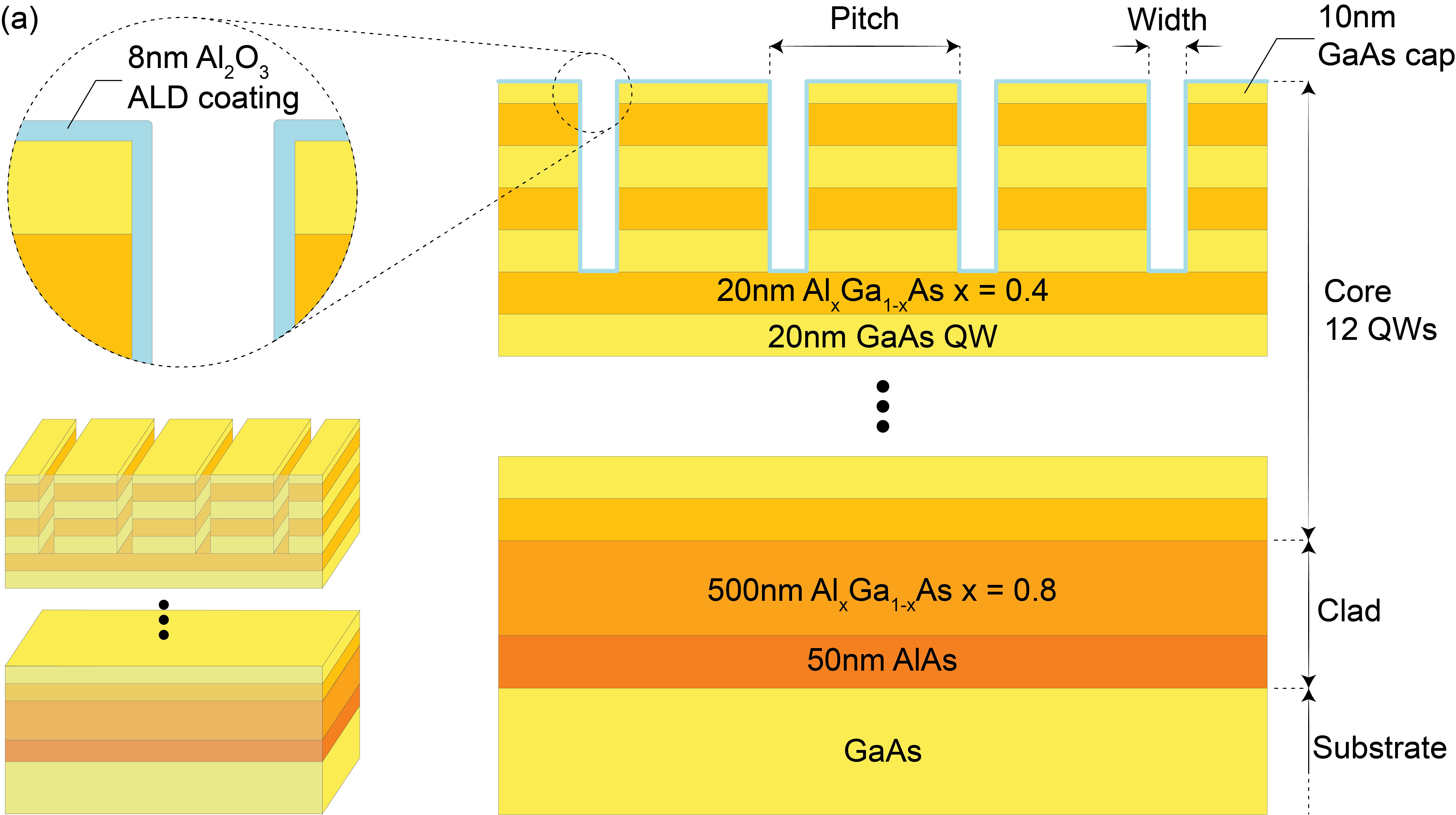}
    \end{subfigure}%
    ~ 
    \hspace{11mm}
    \begin{subfigure}[t]{0.5\textwidth}
        \centering
        \includegraphics[scale=0.36]{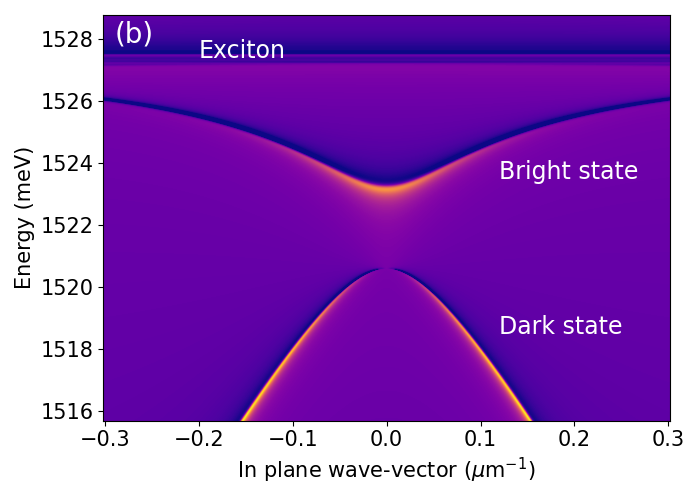}
    \end{subfigure}
    \caption{(a) Sketch of the final processed waveguide sample with a 90nm deep grating, along with a conformal passivating layer of Al$_{2}$O$_{3}$. (b) Simulations of the exciton-polariton dispersion with bright and dark states coupled to the exciton. The BIC appears here as a dark state at the maximum energy point of the dispersion, corresponding to vertical emission (k$_{\parallel}$ = 0) in the far field. }
    \label{fig:sketch}
\end{figure*}

The sample is composed of a GaAs substrate on which a waveguide slab is grown via molecular beam epitaxy. 500nm of Al$_{0.8}$Ga$_{0.2}$As are firstly grown onto the substrate, forming the waveguide cladding. A set of 12 GaAs quantum wells (QWs) and 13 Al$_{0.4}$Ga$_{0.6}$As barriers 20nm thick were grown on top of the cladding, forming the waveguide core. Lastly, 10nm of GaAs were grown on top of the last barrier to act as a cap layer. The optical mode is confined in the waveguide core through total internal reflection, and here it exchanges energy with the QW excitons leading to propagating polaritons. A shallow etched 1D grating (Fig.\ref{fig:sketch}a) can be used to couple propagating and counterpropagating photonic modes via the slab refractive index modulation, hence we can easily introduce new properties by engineering the photonic dispersion. In our system we can exploit such coupling between photonic modes to introduce a symmetry protected bound state in the continuum which appears in the so-called dark state. If we consider E$_{\text{Bright}}$ and E$_{\text{Dark}}$ as the photonic modes originated from the grating coupling and separated by an energy gap, the energy exchange with the QWs exciton can lead to the following system Hamiltonian
\begin{equation}
    H = \left(\begin{matrix}\text{E}_{\text{Bright}} & \frac{\Omega}{2} & 0 & 0 \\ \frac{\Omega}{2} & X & 0 & 0 \\ 0 & 0 & \text{E}_{\text{Dark}} & \frac{\Omega}{2}  \\ 0 & 0 & \frac{\Omega}{2} & X  \end{matrix}\right)
\end{equation}
The photonic bright and dark states are then hybridized with the exciton, hence the diagonalization leads to exciton-polariton eigenstates with a BIC in the dark lower state (Fig.\ref{fig:sketch}b)\cite{Lu2020EngineeringPoints}. Since the gap depends on the coupling strength between the photonic modes, it also depends on the grating's fabrication parameters, such as the etching depth and the filling factor, defined as $\text{FF} = 1-\frac{\text{width}}{\text{pitch}}$ (Fig.\ref{fig:sketch}a). We used Stanford Stratified Structure Solver (S4) \cite{Liu2012SStructures} to simulate the structure in Fig.\ref{fig:sketch}a, from which we obtained the energy dispersion shown in Fig.\ref{fig:sketch}b. The exciton was introduced in the simulations with a Lorentzian resonance in the GaAs dielectric constant, with a Rabi splitting $\Omega = 13.9$meV \cite{Suarez-Forero2021EnhancementInteractions}. The bound state in the continuum appears as a saddle point of the dark state as shown in \cite{Ardizzone2021PolaritonContinuum}, which creates a direct path for polaritons to thermalize from the exciton reservoir. \\
The fabricated gratings are 50$\mu$m wide, 300$\mu$m long with a filling factor 70$\%$ - 75$\%$. The periodicity was set around 240nm to have the photonic mode coupling close to the exciton resonance. We realized the grooves by writing an electron beam sensitive resist, and subsequent pattern transfer into the heterostructure via ICP-Chlorine etching\cite{Liao2017,Maeda1999,Zhou2018,Atlasov2009EffectCrystals}. The etched grating penetrated into the waveguide core, and three different depths were chosen, resulting in different degrees of photonic coupling.

\begin{figure*}[t!]
    \centering
    \begin{subfigure}[t]{0.52\linewidth}
        \centering
        \includegraphics[scale=0.46]{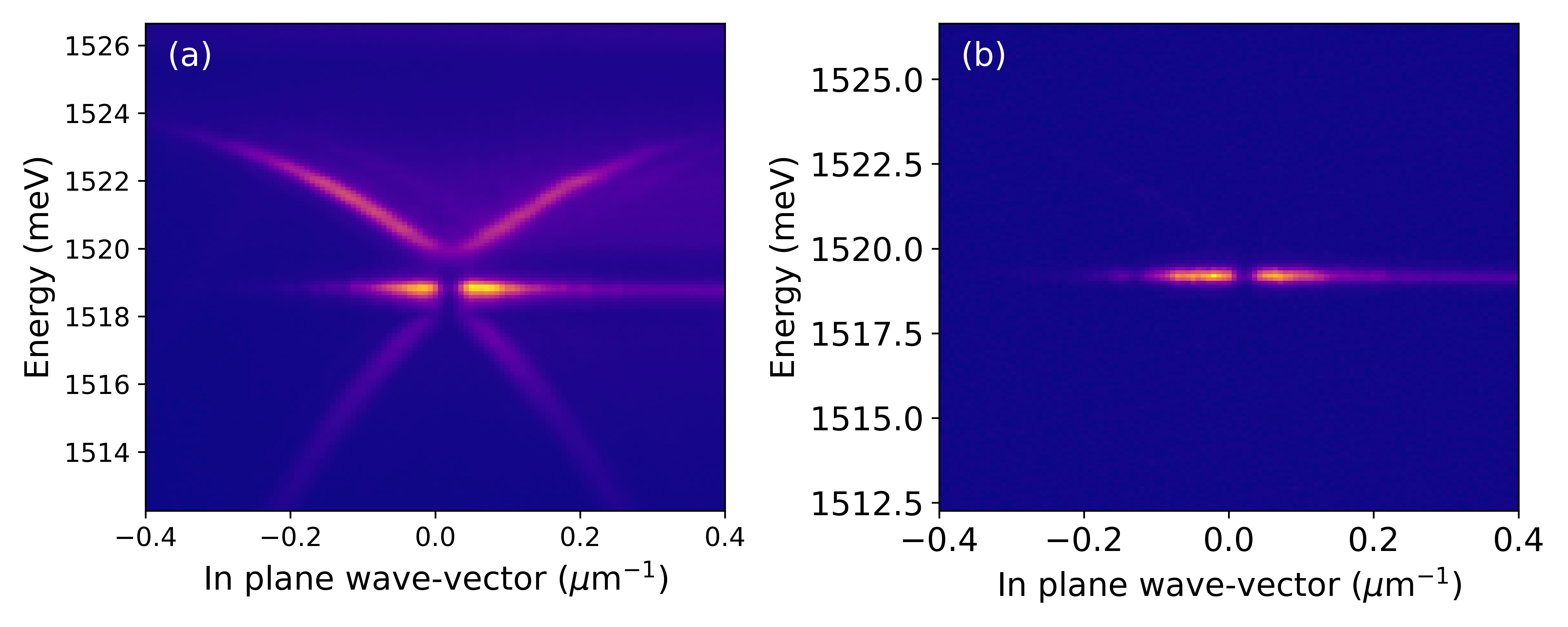}
    \end{subfigure}%
    \begin{subfigure}[t]{0.6\linewidth}
        \centering
        \includegraphics[scale=0.45]{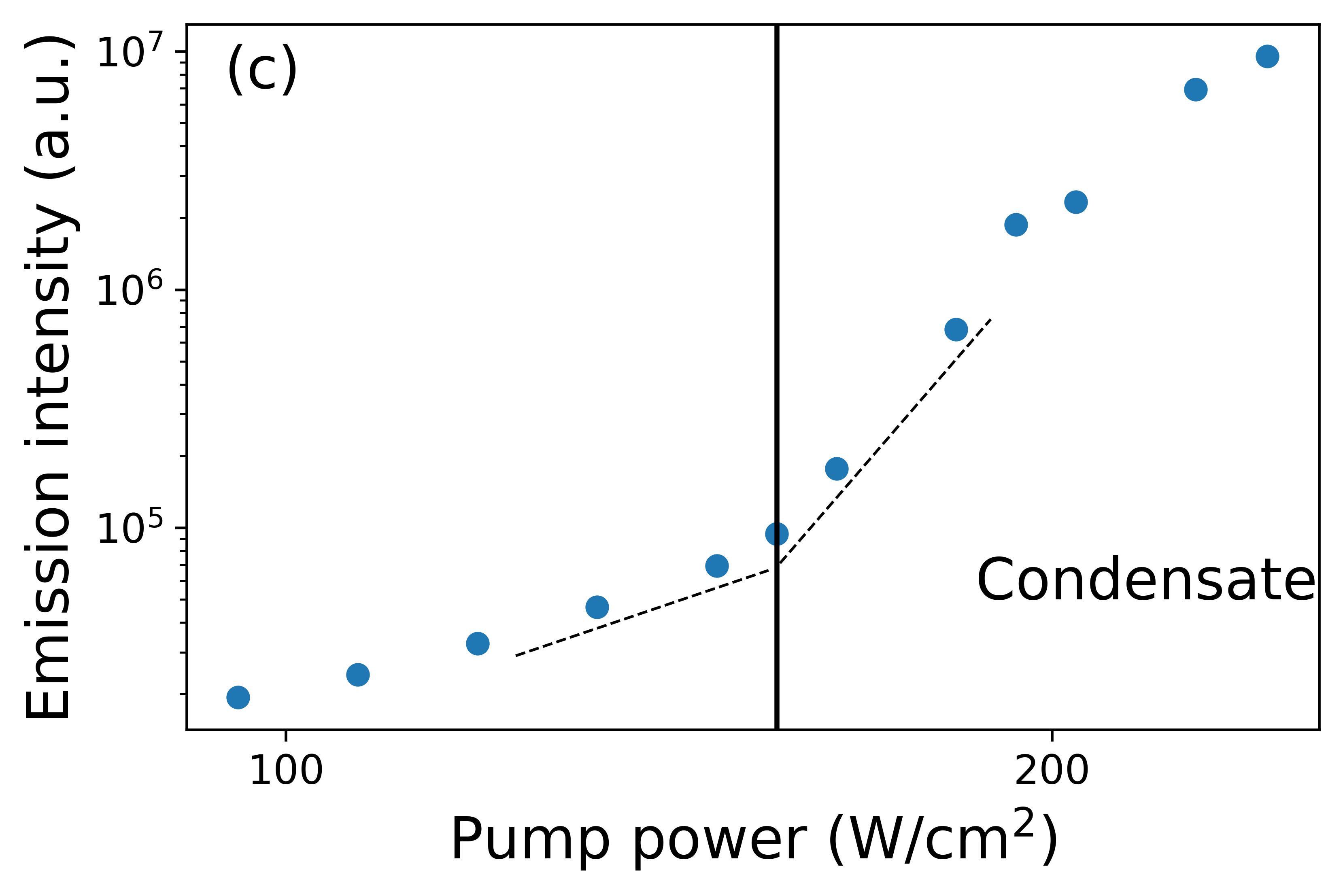}
        \end{subfigure}
    \caption{(a,b) Exciton-polariton energy dispersion and coherent light emission at (a) and above (b) the excitation threshold. The condensate in (b) is blueshifted as a result of polariton-polariton interactions. (c) Emission intensity from the condensate as a function of the pumping power.}
    \label{fig:sketch&dispersion}
\end{figure*}

The sample was cooled down to 4K in a cryostat and polaritons were created by nonresonantly pumping with a laser onto the grating (See Methods). Light was outcoupled by the grating itself and the energy dispersion imaged on a CCD camera. Accumulation of polaritons in the long-lived state triggers the onset of a coherent emission by bosonic stimulation.
The polariton energy dispersion at and above threshold for a grating with filling factor FF = 75$\%$ is shown in Fig.\ref{fig:sketch&dispersion}a and Fig.\ref{fig:sketch&dispersion}b.

\begin{figure}[ht]
\begin{center}
    \includegraphics[scale=0.64]{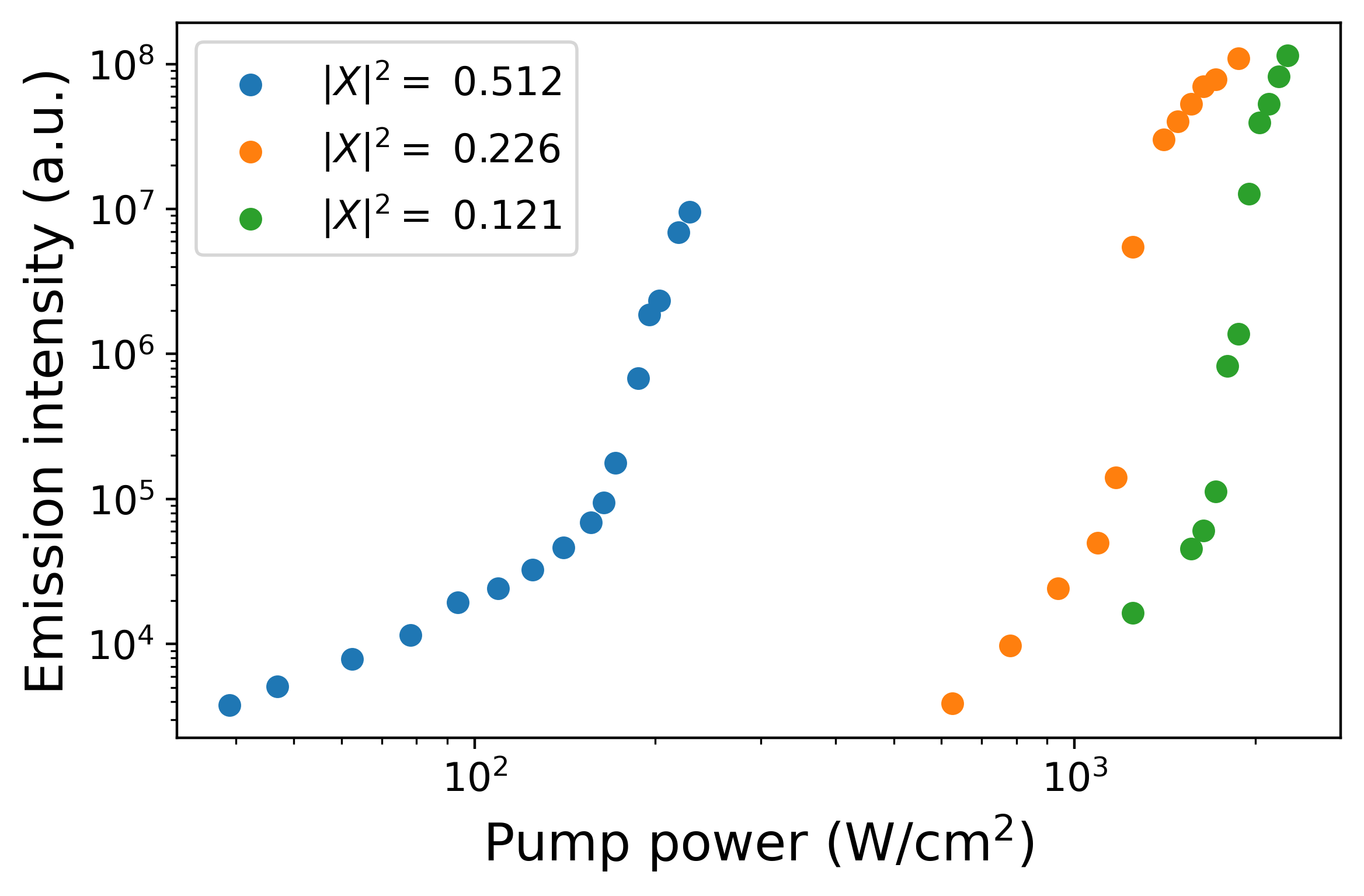}
\caption{Emission intensity from the condensate for periodicities 240nm (blue), 242nm (orange), and 244nm (green). The excitonic fraction $|X|^{2}$ is modified by the grating pitch, favoring a polariton thermalization towards the BIC and reducing the power threshold in the 240nm case.}
\label{fig:thresholdX2}
\end{center}
\end{figure}

\begin{figure}
\centering
\begin{subfigure}[t]{0.50\textwidth}
    \includegraphics[scale = 0.59]{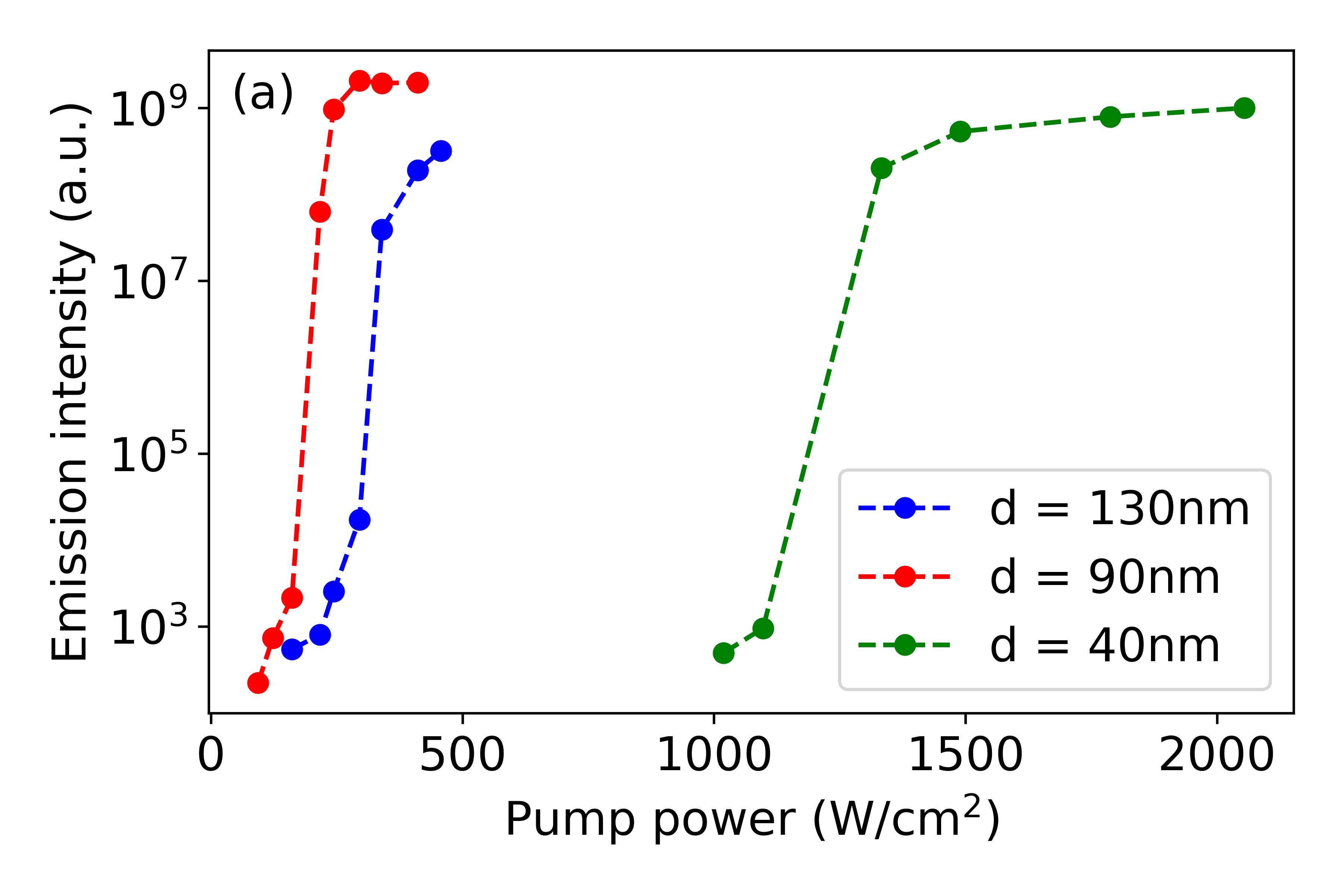}
\end{subfigure}
\begin{subfigure}[t]{0.49\textwidth}
    \centering
    \includegraphics[scale=0.6]{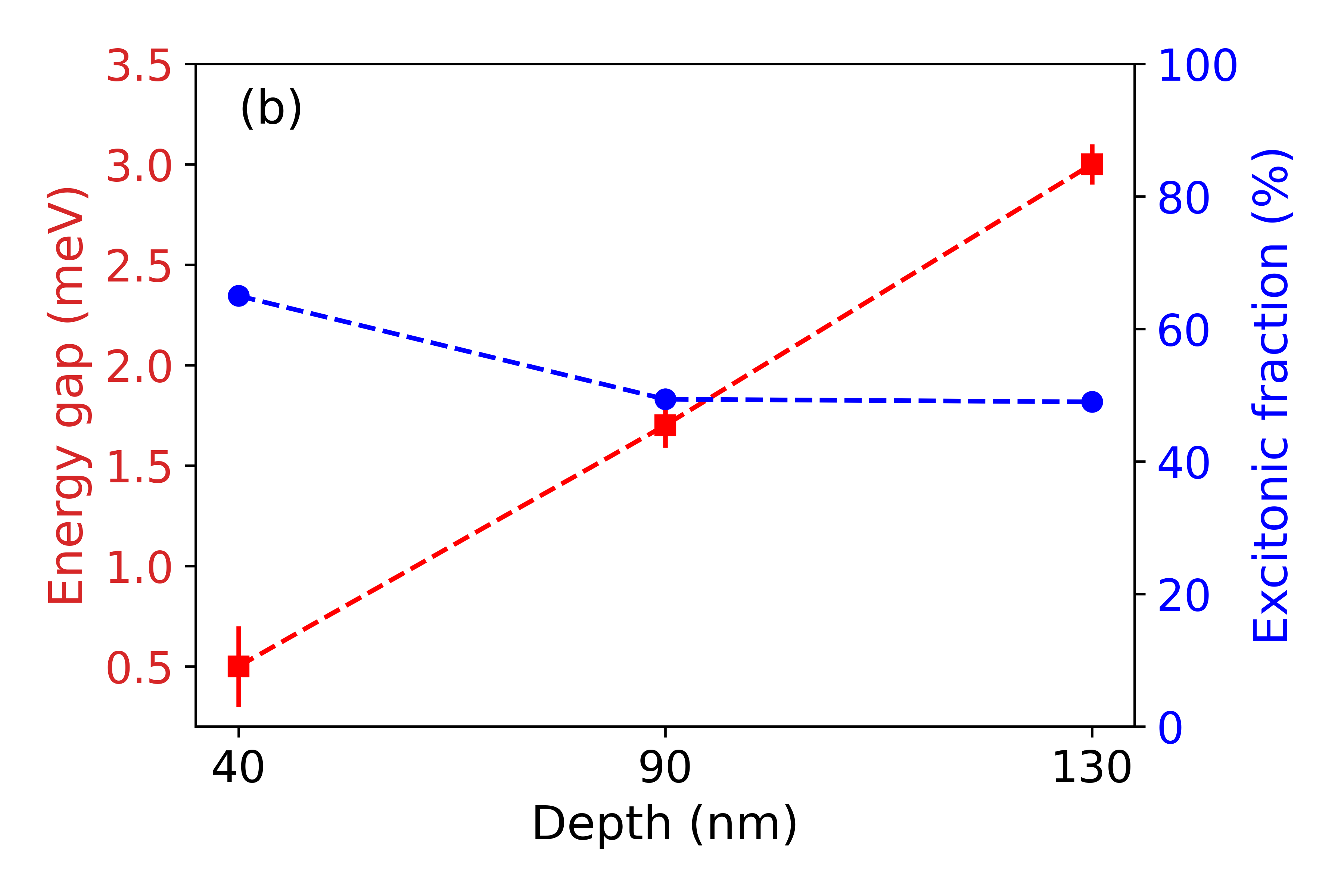}
\end{subfigure}
\caption{(a) Emission intensity as a function of the pump power. The threshold is minimum for a grating 90nm deep, while it is higher for a 40nm deep grating due to smaller gap size, and for a 130nm deep grating due to surface damage. (b) Energy gap and excitonic fraction as a function of the etching depth. The filling factor is approximately the same in all the three cases.}
\label{fig:threshold3depths}
\end{figure}

The power at which the condensation occurs was found to be strongly dependent on the grating parameters. Fig.\ref{fig:thresholdX2} shows the threshold intensity in a 90nm deep grating as a function of the periodicity. We chose gratings with periodicities equal to 240nm, 242nm, and 244nm which correspond to excitonic fractions 51.2$\%$, 22.6$\%$, and 12.1$\%$, respectively. We observed a lower threshold for the 240nm pitch with respect to the 242nm and 244nm cases, ascribable to the higher exciton fraction. In our system, the grating periodicity moves the BIC closer to or further from the exciton, making the polariton thermalization process more or less efficient for the two cases. 
\\
In order to reduce the threshold, different etching depths $d$ were studied with filling factors close to 70$\%$: 40nm, 90nm, 130nm and the measured dispersions are reported in Fig.S6d-f. The etching is performed in the waveguide core, which has consequences on the polariton modes quality. The trap states in the electronic band gap created during the sample's fabrication lead to nonradiative exciton surface recombination at the QWs sidewalls. Surface recombination can be minimized through oxide removal and surface passivation, achievable through etching at sufficiently low ion energy to remain in the ion-assisted chemical etching regime\cite{Glembocki1995EffectsSurface,Leonhardt1998SurfaceArsenide,Eddy1997GalliumProcess}.
For this reason, the sample was immersed in a hydrochloric acid bath to remove the surface oxide, and then 8nm of Al$_{2}$O$_{3}$ were deposited via atomic layer deposition\cite{Guha2017,Hinkle2008,Mikulik2018SurfaceMicropillars,Dhaka2016ProtectiveDeposition} (ALD) at 300°C \cite{Levitskii2018AnnealingHeterostructure} (see Supplementary Material).
The observed emission intensities in the three cases along with the gap size and excitonic fraction are shown in  Fig.\ref{fig:threshold3depths}a,b. The lower threshold observed in deeper gratings is caused by a larger gap. In fact, the BIC state is more protected from the lossy bright state when the gap is larger. In other words, the smaller the gap, the closer the BIC is to the bright state, the higher the losses induced by leakage of the polariton population into the lossy state.
Concurrently with the etching depth, the surface damage increases as demonstrated by the 130nm deep grating in which the condensate was achieved only from one structure with a high filling factor, and only after post-processing, as a consequence of a more pronounced damage (see Fig.S11). The damage induced by the etching was found to act mainly on the excitonic component, as shown in Fig.\ref{fig:thresholdpre&post} and it is discussed in the supplementary material where cathodoluminescence hyperspectral maps elucidate the effect of surface damage on the emission\cite{Negri2020QuantitativeMaterials,Hovington1997CASINO:Program,Drouin2007CASINOUsers}.

\begin{figure}[ht]
  \centering
  \begin{tabular}{@{}c@{}}
    \includegraphics[width=\linewidth]{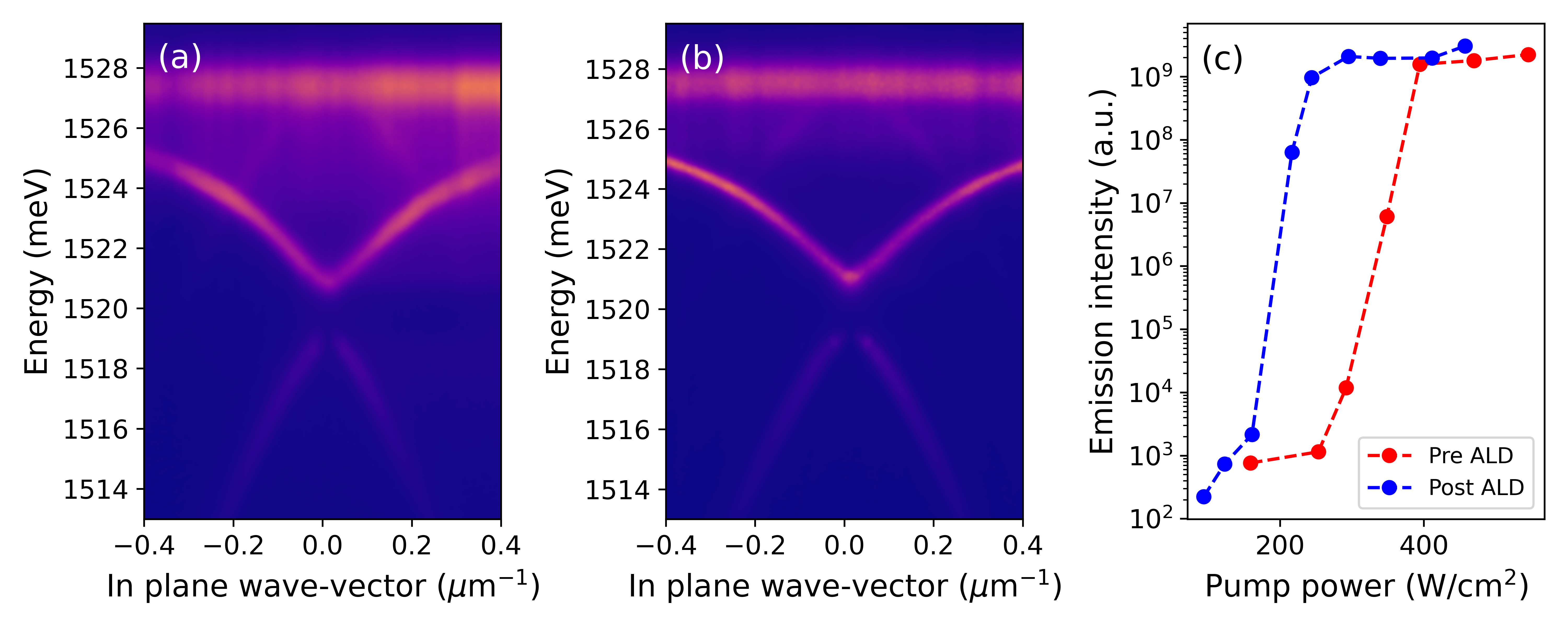} 
  \end{tabular}

  \begin{tabular}{@{}c@{}}
    \includegraphics[scale = 0.65]{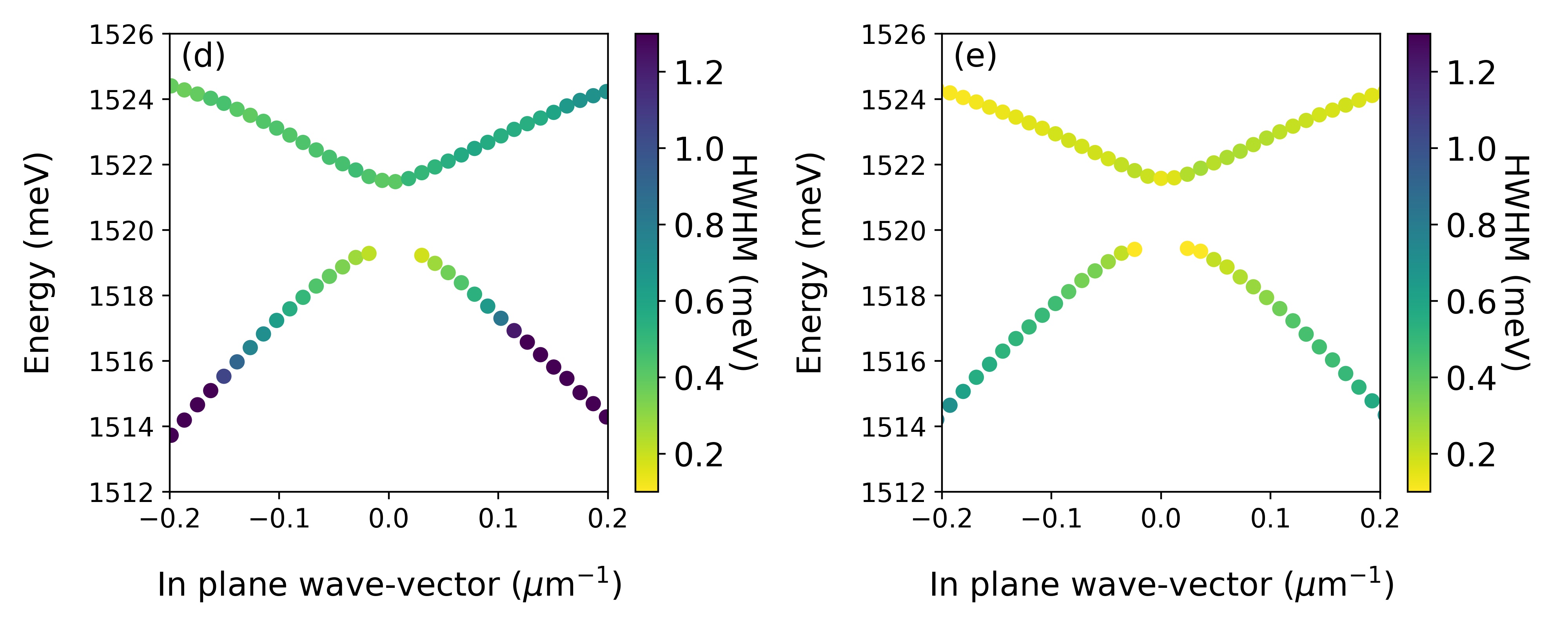}
  \end{tabular}
  \caption{(a,b) Exciton-polariton energy dispersion pre- and post-processing, respectively. (c) Emission intensity at the threshold before and after aluminum oxide deposition for a 90nm deep grating, 240nm pitch. (d,e) Exciton-polariton dispersions corresponding to the case (a) and (b). Colors represent the HWHM fit of the linewidths.}\label{fig:thresholdpre&post}
\end{figure}
Lastly we observed the effect of post-processing on gratings characterized by 240nm pitch and 90nm deep grooves before and after passivation. 
The polariton dispersion for this grating is shown in Fig.\ref{fig:thresholdpre&post}a and Fig.\ref{fig:thresholdpre&post}b, where the threshold reached an energy per pulse corresponding to a few $\frac{\mu \text{J}}{\text{cm}^2}$ after ALD as displayed in Fig.\ref{fig:thresholdpre&post}c. This measurement was performed on several gratings, and all of them showed the same characteristic reduction (See Fig.S10). The longer polariton lifetime, which we associate to lower nonradiative exciton recombination at the groove sidewalls, can be observed in terms of linewidth reduction after Al$_{2}$O$_{3}$ deposition (Fig.\ref{fig:thresholdpre&post}e). This result, even though limited by the spectrometer resolution and inhomogeneous broadening, clearly highlights the reduction of exciton inhomogeneous broadening which ultimately corresponds to a longer exciton-polariton lifetime. A more detailed discussion about the BIC lifetime is presented in \cite{Ardizzone2021PolaritonContinuum} where a lower limit is set by means of polariton propagation under the grating. 
While the effect on the excitonic component is clear, no significant shift in the polariton dispersion was observed after the aluminum oxide deposition, a sign that the lower Al$_{2}$O$_{3}$ refractive index (n = 1.6 for thin film) compared to the photonic mode effective refractive index (n = 3.359) does not alter the designed structure behaviour. 

\section{Conclusions}
The main limitation in achieving thresholdless exciton-polariton lasing is hindered by non-radiative and radiative losses that are inevitably present in microcavities. In this work we have realized a low-density exciton-polariton condensate in a horizontal cavity that makes use of the long lifetime of a quasi-bound state in the continuum in which lifetime is increased by several orders of magnitude due to the special nature of this dark state. However, surface damage during processing introduces nonradiative channels through which excitons can decay. By performing surface ALD post-processing we greatly reduce such an effect, increasing the polariton lifetime at the BIC. Our method substantially lowers the power needed to achieve a polariton condensate. This system is not only a good candidate for low threshold lasers, but also for those experiments that have been affected by the simultaneous presence of horizontal and vertical polariton lasing as recently reported in \cite{Jamadi2019CompetitionMicrocavities}.
In conclusion we assessed the role of grating parameters and fabrication in the achievement of state-of-the-art polariton BEC arising from a bound state in the continuum coupled to an excitonic resonance. The extremely long lifetime of the polariton BIC facilitates the thermalization of the bosonic gas with the onset for condensation at a low record polariton density despite the presence of lower energy states.  Finally we also observed the reduction of the threshold and improvement in the polariton lifetime after hydrochloric acid and passivation of the sidewalls via atomic layer deposition, reducing the density of nonradiative recombination centers at the QW sidewalls. In the case of deep etching, the role of ALD was crucial, given it was able to restore the possibility to observe the condensate from one 130nm deep grating, while it reduced the threshold in all the 90nm deep gratings. The linewidth narrowing observed after post-processing is related to the reduction of non-radiative recombination, improving the overall quality factor of the sample leading to an estimated polariton lifetime of a few hundreds of picoseconds in the BIC\cite{Ardizzone2021PolaritonContinuum}. Since BEC formation from a BIC has the great advantage of reaching the phase transition at extremely low densities, it can be used to obtain ultra-low threshold lasers.
The techniques and ideas investigated and developed here will not only finally boost the realization of polariton-based thresholdless lasers, but could also be a way to realize new horizontal electrically pumped polariton light sources.
\section{Materials and Methods}
\textbf{Optical Measurements}.
The sample was kept at 4K for all the experiments. All measurements were performed using a 150fs, 80MHz pump laser with its central energy tuned at 1.588meV (780nm), and a spot size of 15$\mu$m FWHM. The emission from the sample was sent to a SpectraPro-300i spectrometer with a 1200lines/mm grating in order to reconstruct the reciprocal space. The optical setup allowed the simultaneous acquisition of Fourier space and real space images, which made it possible to keep the sample in focus at all times. To remove spurious effects, the laser tail was cut at around 800nm by a short pass filter, combined with a spatial filter which removes any emission outside the selected grating region.


\textbf{Acknowledgments}
We are grateful to Ronen Rapaport for inspiring discussions and for sharing information about the sample design. The authors acknowledge the project PRIN Interacting Photons in Polariton Circuits INPhoPOL [Ministry of University and Scientific Research (MIUR), Grant No. 2017P9FJBS\_001]. Work at the Molecular Foundry was supported by the Office of Science, Office of Basic Energy Sciences, of the U.S. Department of Energy under Contract No. DE-AC02-05CH11231. We acknowledge the project FISR—C. N. R. Tecnopolo di nanotecnologia e fotonica per la medicina di precisione—CUP B83B17000010001 and ”Progetto Tecnopolo per la Medicina di precisione, Deliberazione della Giunta Regionale Grant No. 2117. This research is funded in part by the Gordon and Betty Moore Foundations EPiQS Initiative, Grant No. GBMF9615 to L. N. Pfeiffer, and by the National Science Foundation MRSEC Grant No. DMR 1420541. L. Francaviglia acknowledges funding from the Swiss National Science Foundation (SNSF) via Early PostDoc Mobility Grant No. P2ELP2\textunderscore184398. Triennal Programm 2021 -2023. Italian Ministry of Research (MIUR) through the FISR 2020 – COVID, project “Sensore elettro-ottico a guida d'onda basato sull'interazione luce-materia” (WaveSense), FISR2020IP\_04324.

\textbf{Disclosures}
The authors declare no conflicts of interest.

\bibliographystyle{apsrev4-1}
\nocite{*}
\input{main.bbl}





\newpage
\textbf{\large{Supplementary Material: Fabrication of Nanostructured GaAs/AlGaAs Waveguide for Low-Density Polariton Condensation from a Bound State in the Continuum}}
\section{Fabrication and post-processing}
\begin{figure}[ht]
\begin{center}
    \includegraphics[width=\linewidth]{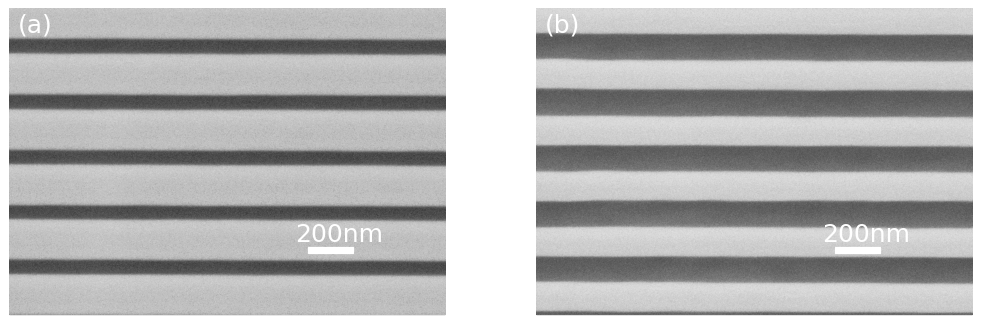}
\caption{Scanning electron microscope images of gratings with filling factors a) 70$\%$ and b) 55$\%$ with darker regions corresponding to the etched grooves.}
\label{fig:SEM}
\end{center}
\end{figure}

\begin{figure}[ht]
\begin{center}
    \includegraphics[scale=1]{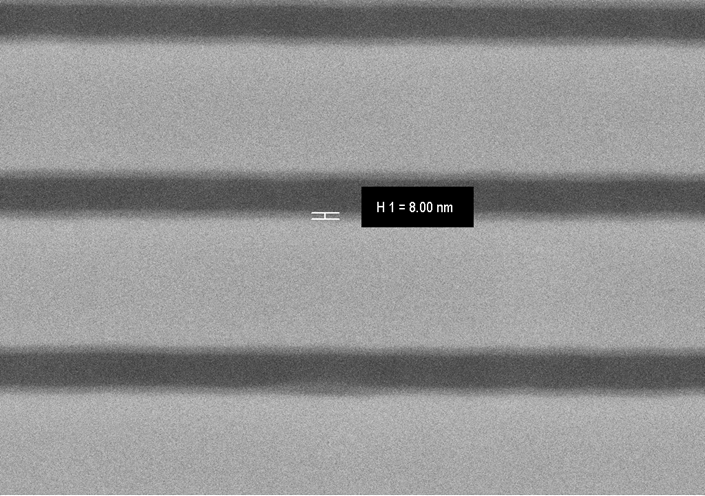}
\caption{SEM image of the 8nm thick aluminum oxide grown via atomic layer deposition. The layer is conformal with the sample surface, hence it covers the grooves sidewalls, bottom and top surface alike.}
\label{fig:SEM_ALD}
\end{center}
\end{figure}

The heterostructure  was  grown  via  molecular beam  epitaxy on a GaAs substrate. The waveguide cladding is made of Al$_{0.8}$Ga$_{0.2}$As 500nm thick, while the core is composed of 13 barriers of Al$_{0.4}$Ga$_{0.2}$As barrier (20nm thick), 12 GaAs QWs (20nm thick), and a final cap layer made of GaAs (10nm thick). We then fabricated a 1D grating by etching the waveguide, inducing a coupling between propagating and counter-propagating modes. The grating period was tuned to have the coupled photonic states at energies close to the exciton.\\
An electron beam sensitive resist (ZEP520a diluted 50$\%$ in anisole) was spun onto the heterostructure and baked at 180°C for 5 minutes, leading to 160nm thick final resist layer. The resist was exposed with a Vistec VB300 Electron Beam Lithography System, with doses that range from 140$\mu$C/cm$^{2}$ up to 280$\mu$C/cm$^{2}$. The electron beam dose and design allow to achieve a wide variety of filling factors (see Fig.\ref{fig:SEM}) that range from 30$\%$ to 75$\%$, necessary to tune the BIC and band gap. Each grating has a dimension of 300$\mu$m x 50$\mu$m. The resist was then developed in amyl acetate, and the heterostructure was etched using Oxford PlasmaLab 150 Inductively Coupled Etcher. A recipe with Cl$_{2}$:BCl$_{3}$:N$_{2}$=11:3:11\cite{Atlasov2009EffectCrystals} was found to be ideal for low selectivity between GaAs and Al$_{0.4}$Ga$_{0.6}$As, while guaranteeing vertical sidewalls and good selectivity with the softmask ($\simeq $ 2.73). In the past, the damage induced by energetic ions below 200eV was proved to be caused by an ion-assisted chemical etching which can be recovered through passivation\cite{Leonhardt1998SurfaceArsenide}. For this reason, the RF and IPC power were also tuned to reduce the ion energy down to 138eV in order to limit the etching induced damage in the GaAs QWs. \\
The GaAs oxide at the surface was removed via hydrochloric acid and was prevented to be formed again by passivating the surface with 8nm of Al$_{2}$O$_{3}$ conformal layer right after the acid step. The sample was placed in a Oxford FlexAl-Plasma Enhanced Atomic Layer Deposition and kept for 10 minutes at a temperature of 300°C before starting the deposition\cite{Levitskii2018AnnealingHeterostructure}. The alumina layer was conformally grown all over the sample to reduce the nonradiative recombination rate at the GaAs grating sidewalls\cite{Guha2017,Dhaka2016ProtectiveDeposition,Mikulik2018SurfaceMicropillars}. The SEM image in Fig.\ref{fig:SEM_ALD} shows the top view of the grating with the thin film at the etched grooves. The results shown in Fig.5d and Fig.5e of the main text show the linewidth fit of the exciton-polariton dispersions before and after the passivation (Fig.5a and Fig.5b, respectively). The reduction of the exciton linewidth led to a quality improvement of the sample after the passivation with the consequent reduction of the minimum polariton density needed to achieve condensation. 

\section{Simulations}
\begin{figure}[ht]
\centering
\begin{subfigure}[t]{0.49\textwidth}
    \includegraphics[scale = 0.4]{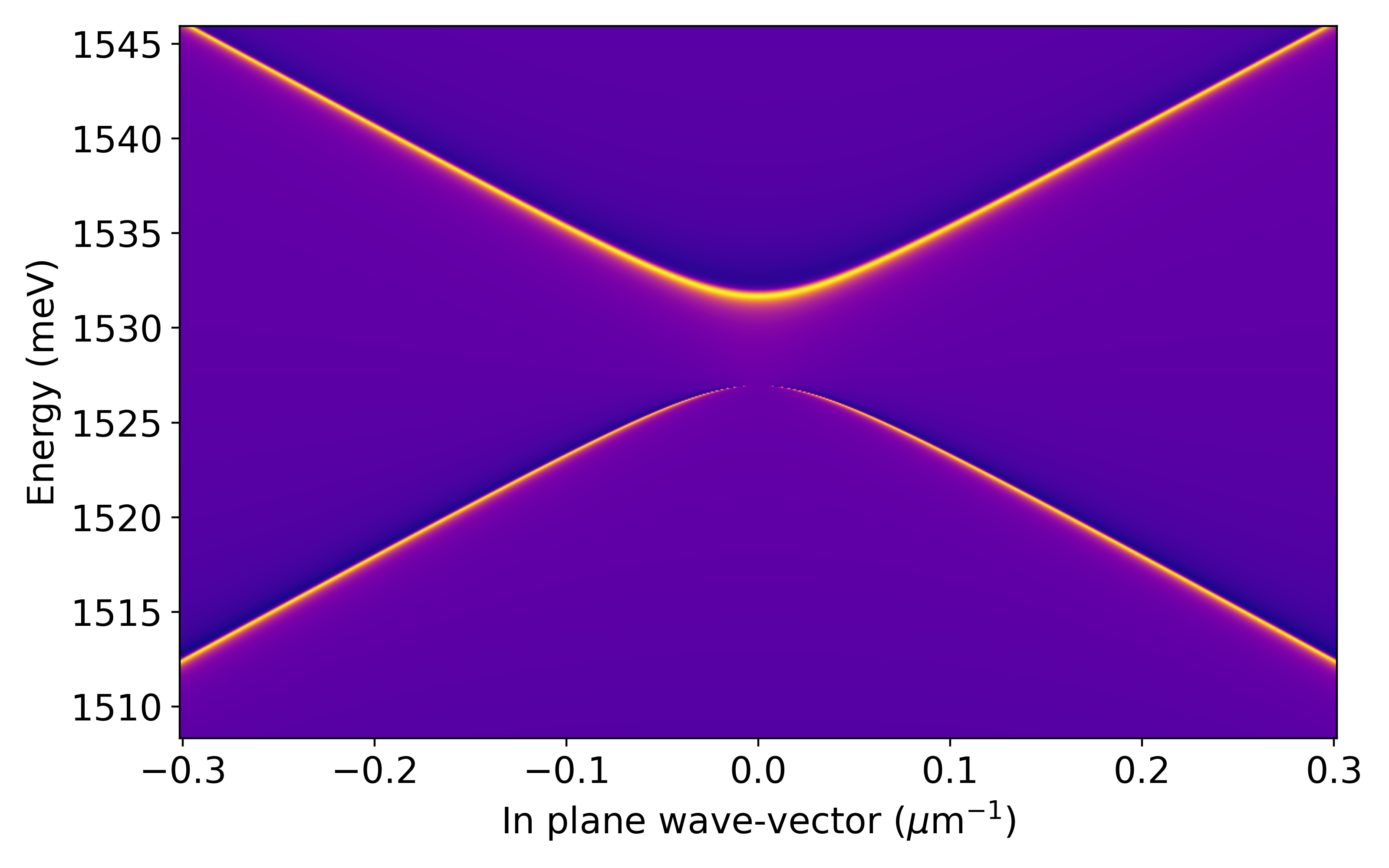}
    \caption{}
\end{subfigure}
\begin{subfigure}[t]{0.49\textwidth}
    \centering
    \includegraphics[scale=0.4]{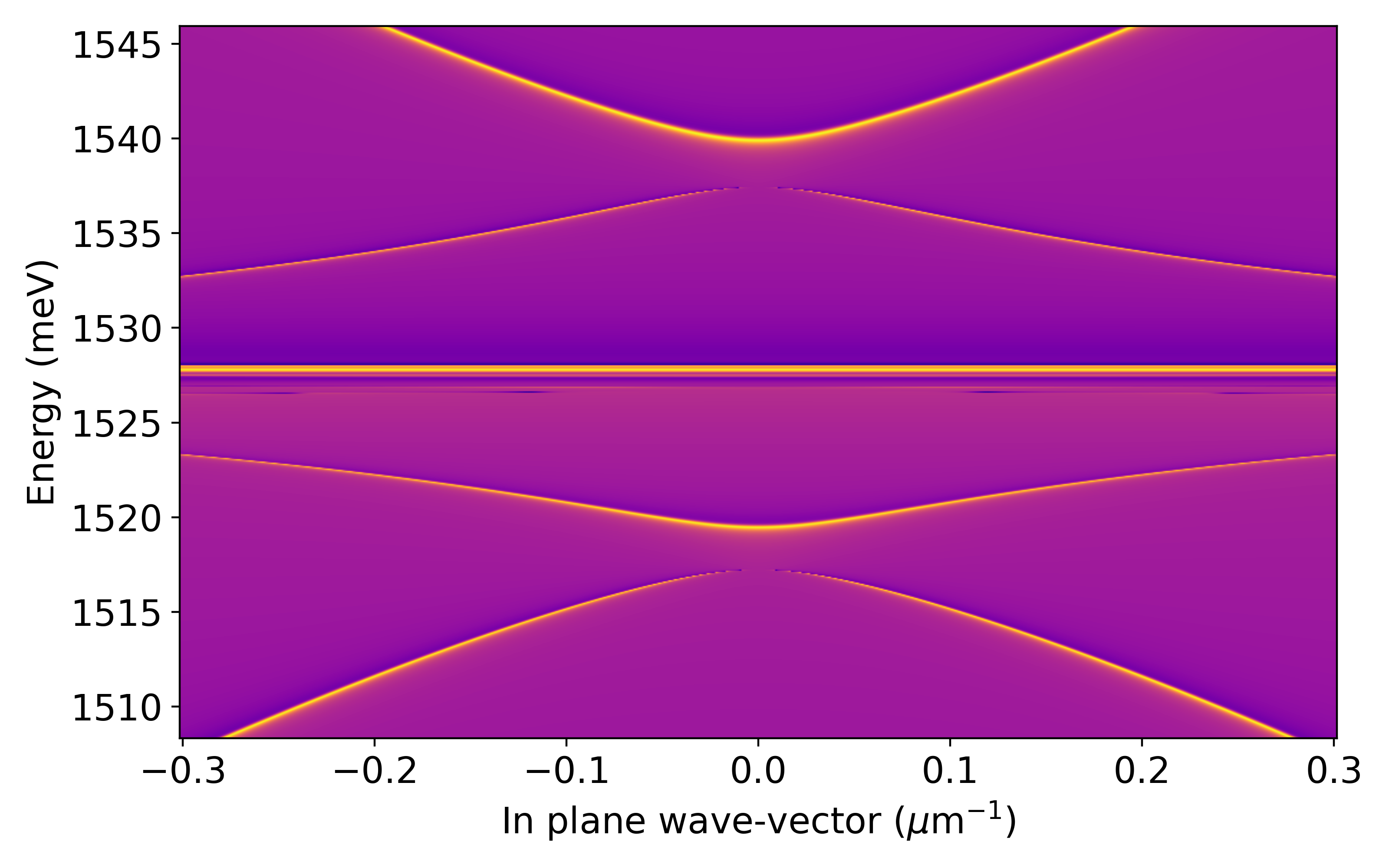}
    \caption{}
\end{subfigure}
\caption{(a) Energy dispersion of the coupled photonic modes. (b) Upper and lower exciton-polariton dispersion branches that arise from the coupling between the two photonic modes in (a) with the exciton.}
\label{fig:simulationsPhEx}
\end{figure}

Simulations of the grating were performed using Stanford Stratified Structure Solver (S4). The strong coupling is introduced by means of a modulation of the GaAs dielectric function\cite{article} as a function of the energy E  
\begin{equation}
    \epsilon(E) = \epsilon_{\text{GaAs}} +\epsilon_{\text{GaAs}} \frac{\omega_{LT}}{X-E-i\gamma}
\end{equation}
where $\epsilon_{\text{GaAs}} = 3.54$ is the GaAs dielectric constant at 10K, $X = 1527.5$ meV and $\gamma = 0.01$ meV are the exciton resonance and its losses, respectively. $\omega_{LT}$ is the longitudinal transverse splitting which, in close proximity to the resonance frequency, is defined as
\begin{equation}
    \omega_{LT} = \frac{\Omega^{2}}{2X}
\end{equation}
where $\Omega = 13.9$ meV is the Rabi splitting calculated from the dispersion fitting in \cite{Suarez-Forero2021EnhancementInteractions,Rosenberg2018}. The barrier dielectric constant was set to $\epsilon_{\text{Al}_{0.4}\text{Ga}_{0.6}\text{As}} = 3.3$. While the bound state in the continuum is introduced through the coupling of pure photonic modes, the presence of the exciton does not induce a finite linewidth as the state is symmetry protected. Nevertheless, its lifetime is still limited by the intrinsic losses inside the waveguide. Fig. \ref{fig:simulationsPhEx}.a shows the photonic modes without the resonance, which is then introduced in Fig. \ref{fig:simulationsPhEx}.b leading to the 4 polariton branches: two above and two below the exciton energy. For the rest of the discussion we will focus on the lower energy branches. 

\begin{figure}[ht]
\centering
\begin{subfigure}[t]{0.32\textwidth}
    \includegraphics[scale = 0.9]{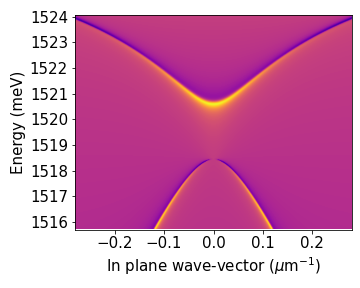}
    \caption{}
\end{subfigure}
\begin{subfigure}[t]{0.32\textwidth}
    \centering
    \includegraphics[scale=0.9]{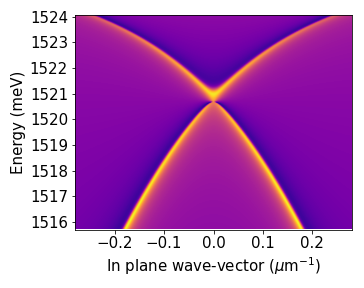}
    \caption{}
\end{subfigure}
\begin{subfigure}[t]{0.32\textwidth}
    \centering
    \includegraphics[scale=0.9]{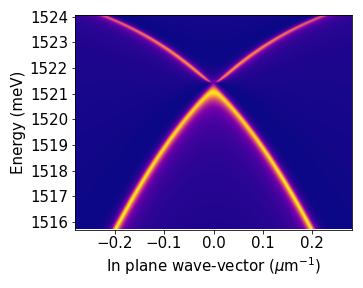}
    \caption{}
\end{subfigure}
\caption{Simulations of the sample for three different filling factors: (a) 75$\%$, (b) 50$\%$, and (c) 25$\%$.}
\label{fig:simulations}
\end{figure}

As the filling factor is reduced from 75$\%$ to 50$\%$ (Fig.\ref{fig:simulations}a,b.), the gap is reduced with a blueshifted dark state. This was also confirmed experimentally as reported in Fig.\ref{fig:gap_exFrac_FF}a. An interesting aspect is the switch of the BIC from the lower state to the upper one (Fig.\ref{fig:simulations}c.), which was also predicted from previous theoretical calculations\cite{Ardizzone2021PolaritonContinuum}. We used S4 to obtain the exciton-polariton dispersion with its geometrical parameter dependence. However, in order to fit our experiments we relied on a coupled oscillator model, described in the next section, which also provides us with the corresponding states excitonic fractions.

\section{Coupled oscillators model}
As reported by Ardizzone et al.\cite{Ardizzone2021PolaritonContinuum}, this system can be described as a 2-steps perturbation method. Using this method, they have developed a theory that describes the polariton states along both k$_{x}$ and k$_{y}$ directions. The first step is to introduce a coupling between the transverse electric propagating ($\hat{\omega}_{+1}$) and counter-propagating ($\hat{\omega}_{-1}$) modes. This is done by introducing a refractive index modulation in our waveguide, while the uncoupled modes were obtained from the fitted dispersion in \cite{Suarez-Forero2021EnhancementInteractions}. Here we show the result of the first step where the photonic modes are coupled with a strength U, limiting our considerations to k$_{y}$=0. Such term can be changed experimentally by varying both filling factors and the depth of the grooves. The Hamiltonian can then be written as

\begingroup
\large
\begin{equation}
    H = \left(
    \begin{matrix}
    \hat{\omega}_{+1} & \text{U}\\ \text{U} & \hat{\omega}_{-1}
    \end{matrix}
    \right) + i \left(
    \begin{matrix}
    \gamma_{r} & -\gamma_{r}\\ -\gamma_{r} & \gamma_{r}
    \end{matrix}
    \right)
\end{equation}
\endgroup\\
where $\gamma_{r}$ represents the radiative losses and $\hat{\omega}_{\pm1}$ are the photonic dispersion characteristics obtained through the fitting of the uncoupled modes\cite{Suarez-Forero2021EnhancementInteractions}.
The eigenvalues for the pure photonic coupled modes are
\begin{equation}
    E_{\text{Bright}} = \frac{\hat{\omega}_{+1}+\hat{\omega}_{-1}}{2} + i\gamma_{r} +\frac{1}{2}\sqrt{(\hat{\omega}_{+1}-\hat{\omega}_{-1})^{2}+4(U+i\gamma_{r})^{2}}
\end{equation}
\begin{equation}
    E_{\text{Dark}} = \frac{\hat{\omega}_{+1}+\hat{\omega}_{-1}}{2} + i\gamma_{r} -\frac{1}{2}\sqrt{(\hat{\omega}_{+1}-\hat{\omega}_{-1})^{2}+4(U+i\gamma_{r})^{2}}
\end{equation}
The second step consists in the coupling of the resulting states with the exciton resonance which eigenvalues can be found by solving the 4x4 Hamiltonian
\begin{equation}
    H = \left(\begin{matrix}\text{E}_{\text{Bright}} & \frac{\Omega}{2} & 0 & 0 \\ \frac{\Omega}{2} & X & 0 & 0 \\ 0 & 0 & \text{E}_{\text{Dark}} & \frac{\Omega}{2}  \\ 0 & 0 & \frac{\Omega}{2} & X  \end{matrix}\right)
\end{equation}
The matrix diagonalization leads to the E$_{\text{dark}}^{\text{pol}}$ and E$_{\text{bright}}^{\text{pol}}$ which describe our polaritonic system
\begin{equation}
    E_{\text{Bright},\pm}^{\text{pol}} = \frac{E_{\text{Bright}}+X}{2} \pm\frac{1}{2}\sqrt{(E_{\text{Bright}}-X)^{2}+\Omega^{2}}
\end{equation}
\begin{equation}
    E_{\text{Dark},\pm}^{\text{pol}} = \frac{E_{\text{Dark}}+X}{2} \pm\frac{1}{2}\sqrt{(E_{\text{Dark}}-X)^{2}+\Omega^{2}}
\end{equation}
The two separate solutions identified by $\pm$ for both bright and dark states correspond to the lower and upper bright and dark states, respectively. These states are also shown in Fig. \ref{fig:simulationsPhEx}.b as upper and lower polariton branches. From now on, we will consider only the lower branches, hence the states below the exciton energy $E_{\text{Dark},-}^{\text{pol}}$ and $E_{\text{Bright},-}^{\text{pol}}$. This model will also be used to determine the excitonic fractions for the rest of the work.
This simple model, even though limited by the lack of explicit definition of the grating coupling strength U, it can qualitatively describe our system. One of the things mentioned and shown in Fig.4b is the experimental trend of the energy gap and excitonic fraction as a function of the etching depth. Here we show that increasing the coupling U between the pure photonic modes can lead to a similar behaviour illustrated in Fig.4b where gratings with the same filling factors were compared when the etching depth was varied. The dispersions obtained from the model for three different coupling factors are shown along with the fitted dispersions for the different etching depths considered. Results are shown in Fig.\ref{fig:dispU}.

\begin{figure}[ht]
\begin{center}
    \includegraphics[scale = 0.8]{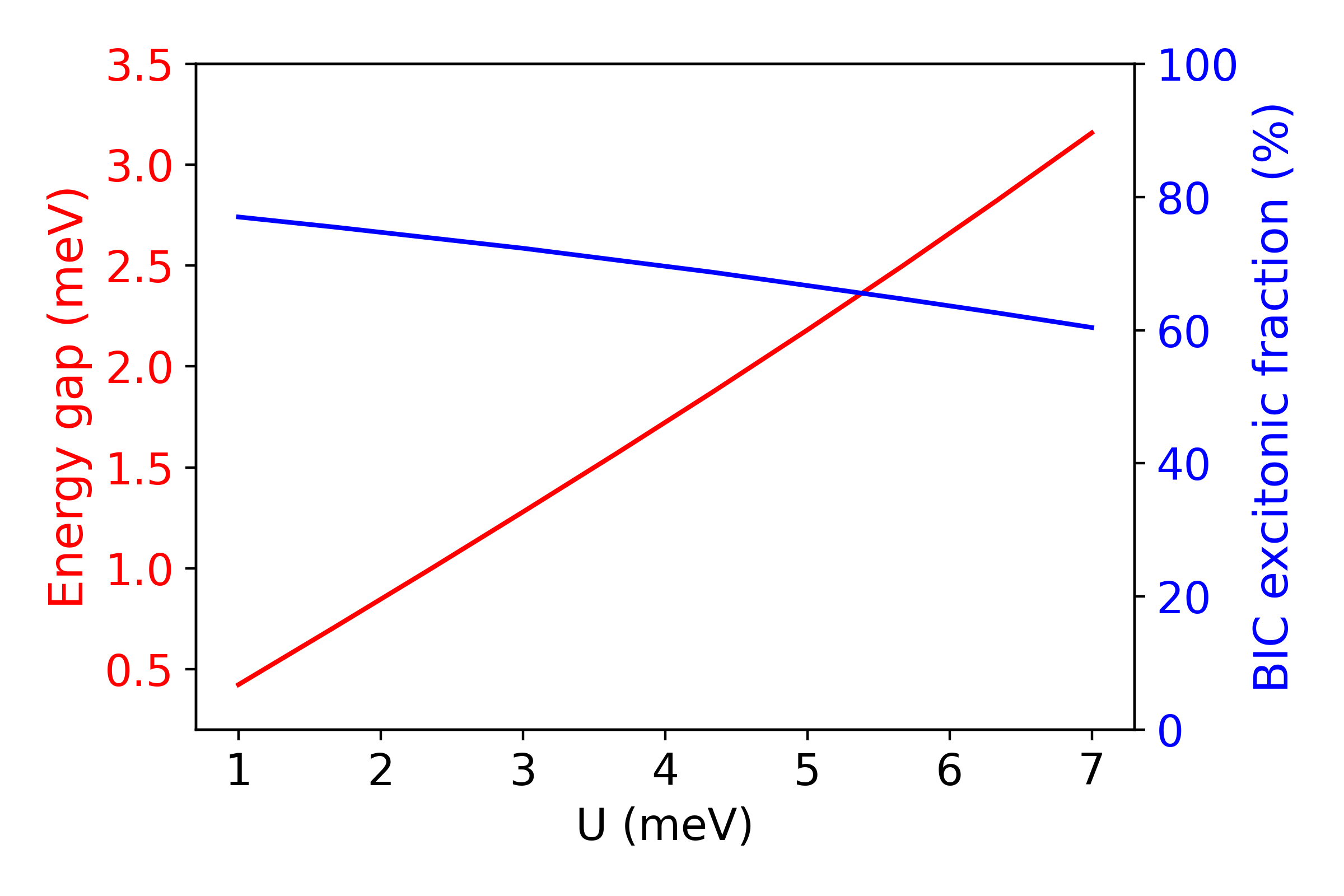}
\caption{Energy gap opened in the polaritonic dispersion along with the excitonic fraction of the BIC obtained through the coupled oscillators model as a function of the grating coupling strength U.}
\label{fig:ExFracTheory}
\end{center}
\end{figure}

\begin{figure}[ht]
\begin{center}
    \includegraphics[scale = 0.85]{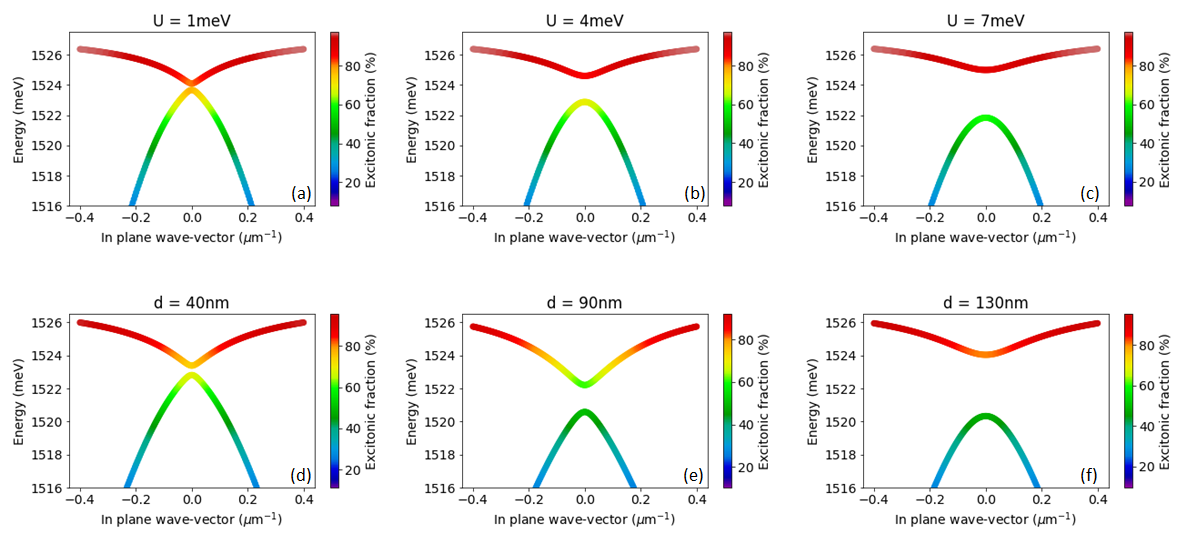}
\caption{(a-c) Theoretical polariton dispersions extracted from the two coupled oscillators model for different coupling coefficients U. (d-f) Fit of the measured dispersions for the etching depths $d$ considered in this work. The colors represent the excitonic fraction.}
\label{fig:dispU}
\end{center}
\end{figure}
\newpage
\section{Control of the polariton dispersion via processing}
\begin{figure}[ht]
\begin{center}
    \includegraphics[scale = 0.68]{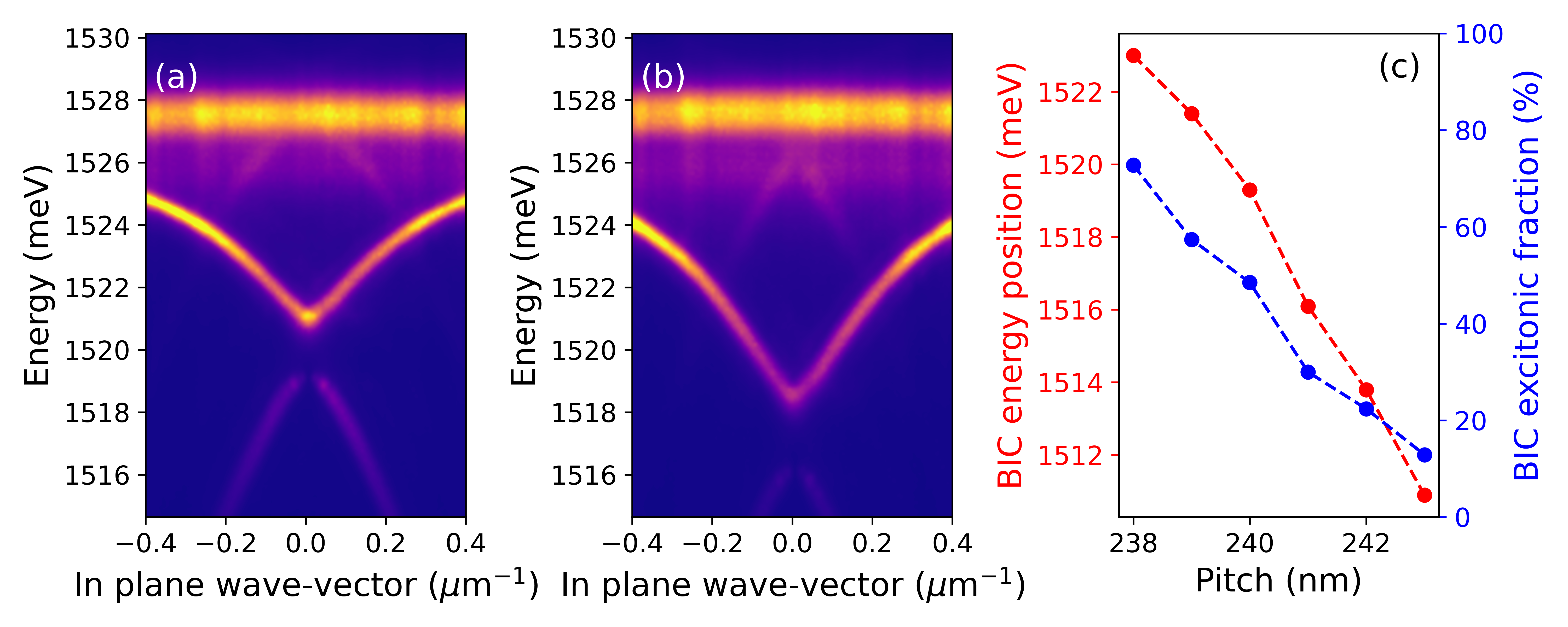}
\caption{(a,b) Dispersions for a pitch equal to 240 nm and 241 nm. (c) BIC energy position (red) and excitonic fraction (blue) as a function of the grating periodicity.}
\label{fig:multiplepitch}
\end{center}
\end{figure}

As mentioned in the main text, the BIC energy, its excitonic fraction, and the gap are parameters that can be tuned by changing the grating periodicity which is a peculiar advantage of our horizontal configuration. In Fig.\ref{fig:multiplepitch} we show such control over the dispersions for pitch equal to 240 nm (a) and 241 nm (b). Along with the tuning of the polariton dispersion, the grating pitch modifies the BIC energy position and excitonic fraction (c). 

\begin{figure}[ht]
\centering
\begin{subfigure}[t]{0.472\textwidth}
    \includegraphics[scale = 0.57]{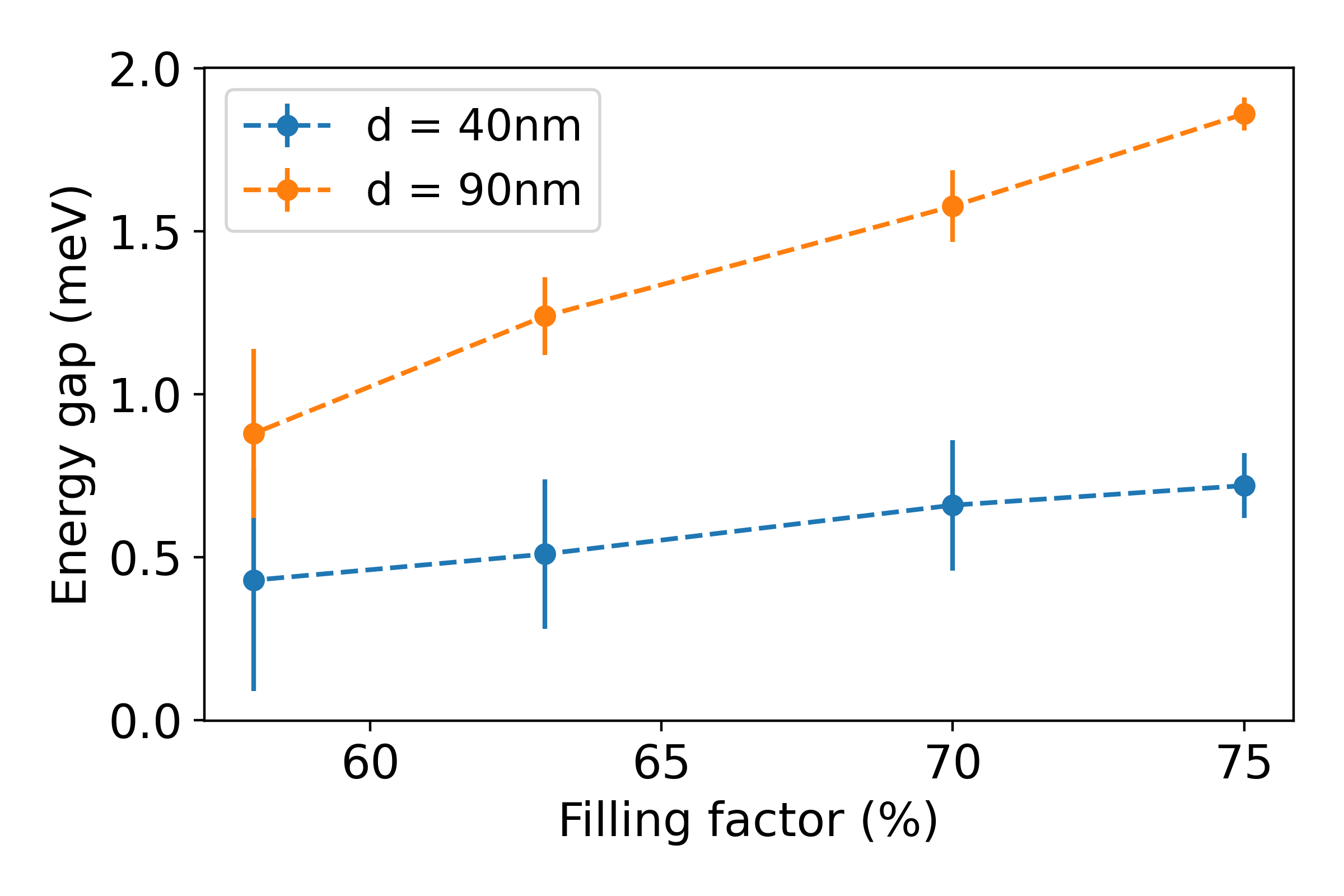}
    \caption{}
\end{subfigure}
\begin{subfigure}[t]{0.52\textwidth}
    \centering
    \includegraphics[scale=0.56]{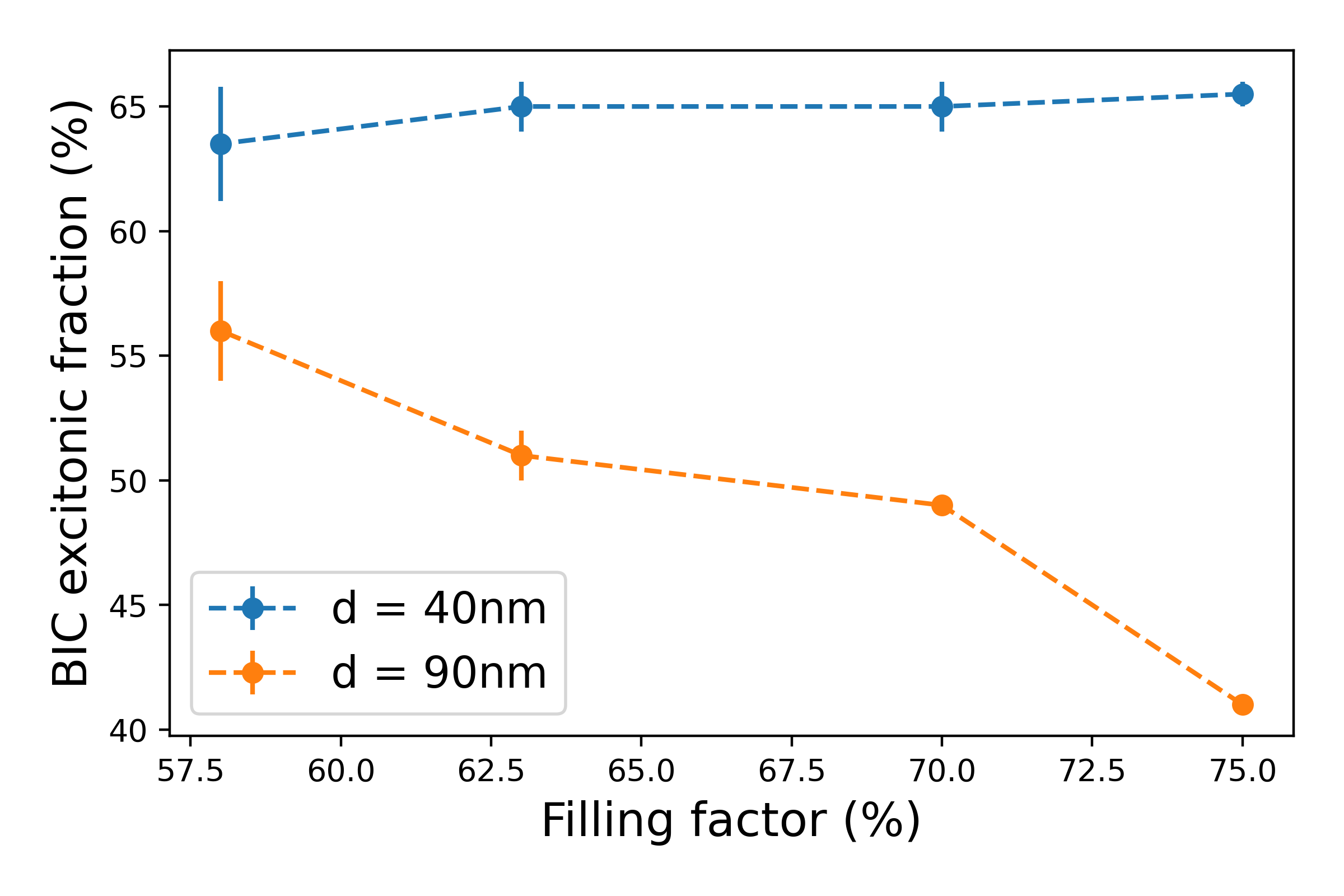}
    \caption{}
\end{subfigure}
\caption{(a,b) Energy gap and excitonic fraction as a function of the filling factor for the 40nm and 90nm deep gratings.}
\label{fig:gap_exFrac_FF}
\end{figure}

The energy gap is another controllable parameter which is modified by the coupling between propagating and counter-propagating mode. While it is almost constant for a 40 nm deep grating, its size is changed with the filling factor in deeper gratings(Fig.\ref{fig:gap_exFrac_FF}a), allowing a further control over the BIC energy position and its excitonic fraction (Fig.\ref{fig:gap_exFrac_FF}b). Fig.\ref{fig:gap_exFrac_FF} shows the energy gap and excitonic fraction obtained through a coupled oscillator model fitting of the polariton dispersions. 

\begin{figure}[ht]
\begin{center}
    \includegraphics[scale = 0.7]{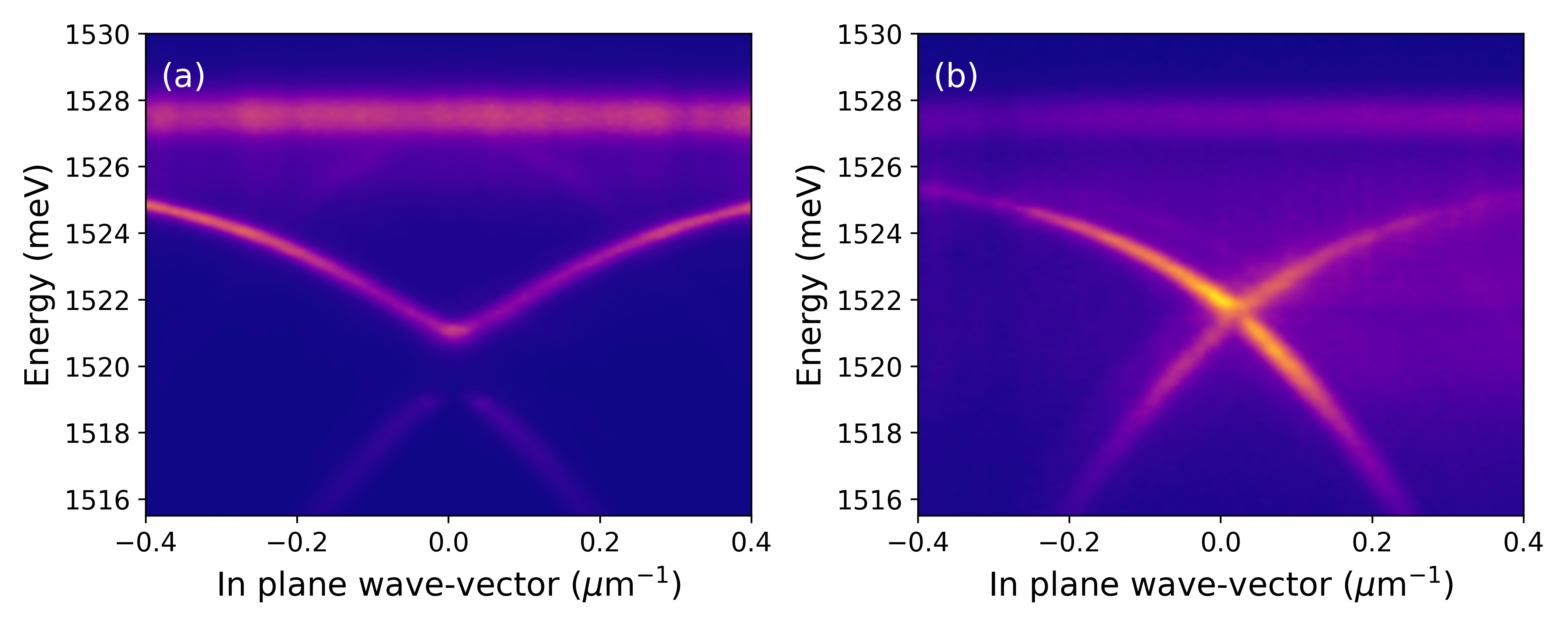}
\caption{Polariton energy dispersion for (a) FF = 75$\%$ and (b) FF = 50$\%$. Passing from (a) to (b), the bright exciton-polariton state remains mostly unaltered while the dark exciton-polariton state blueshifts. The gap closes towards FF$\simeq$ 50$\%$ in which case no lasing was observed. }
\label{fig:multipleFF}
\end{center}
\end{figure}

The data in Fig.\ref{fig:multipleFF}  show the situation for two different filling factors where the gap closes for values close to 50$\%$. No lasing was observed for such gratings as a consequence of higher losses due to a BIC closer to the bright dispersion branch.

\section{Condensation after Atomic Layer Deposition}
\begin{figure}[ht]
\begin{center}
    \includegraphics[scale = 0.7]{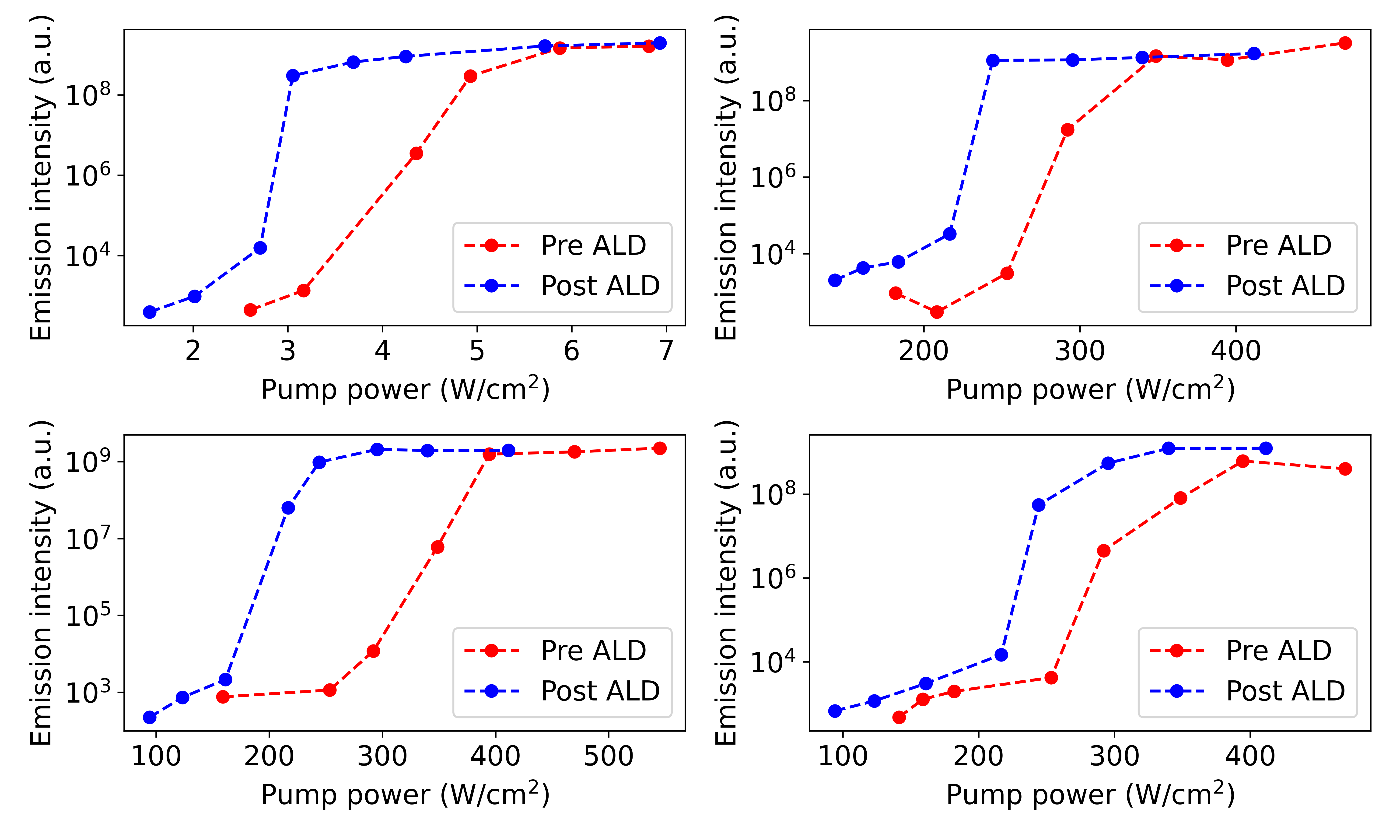}
\caption{Lasing intensity of 4 pre- and post-processed gratings.}
\label{fig:prepostALD_4gratings}
\end{center}
\end{figure}

\begin{figure}
\centering
\begin{subfigure}[t]{0.3\textwidth}
    \includegraphics[scale = 0.36]{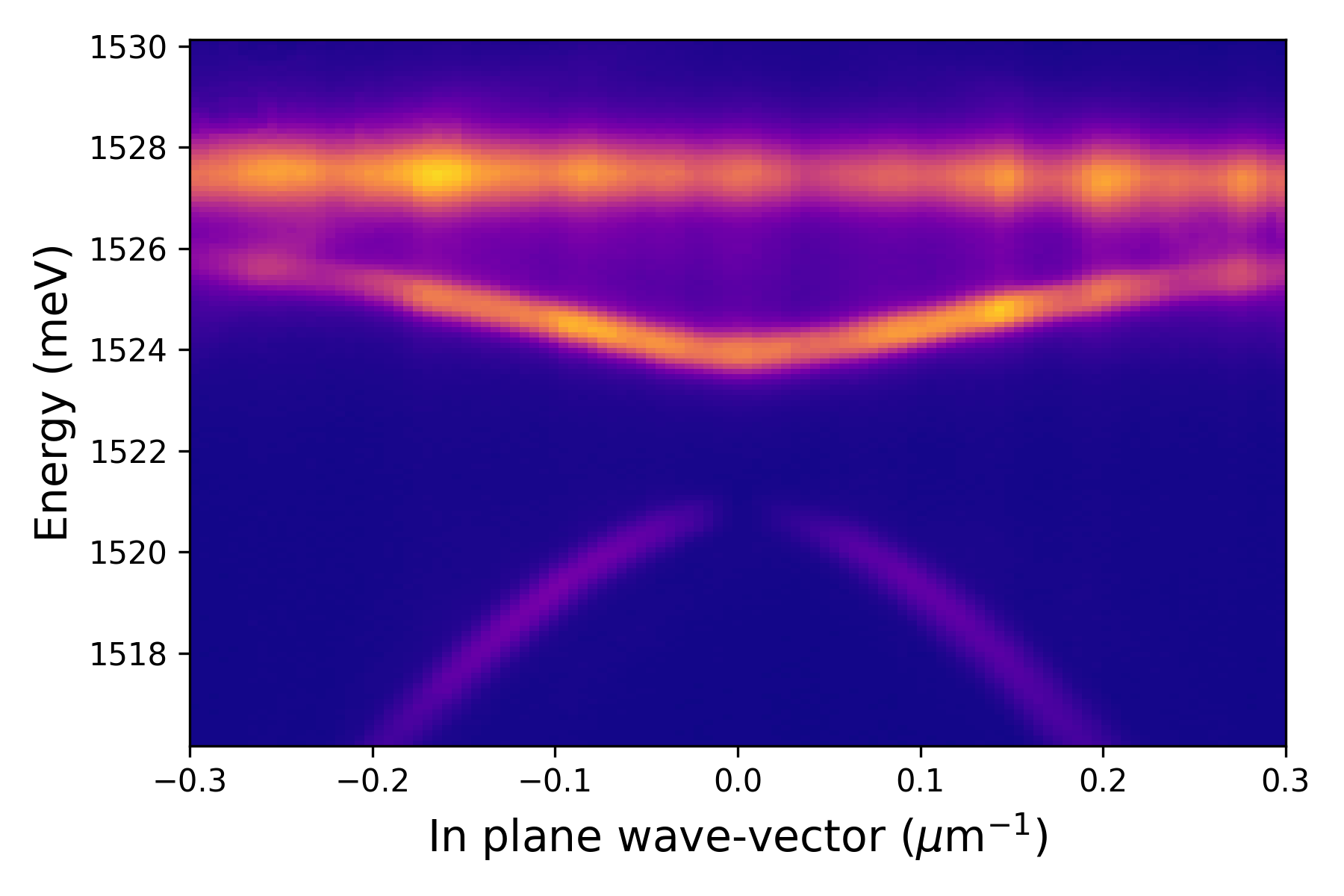}
    \caption{}
\end{subfigure}
\begin{subfigure}[t]{0.3\textwidth}
    \includegraphics[scale = 0.36]{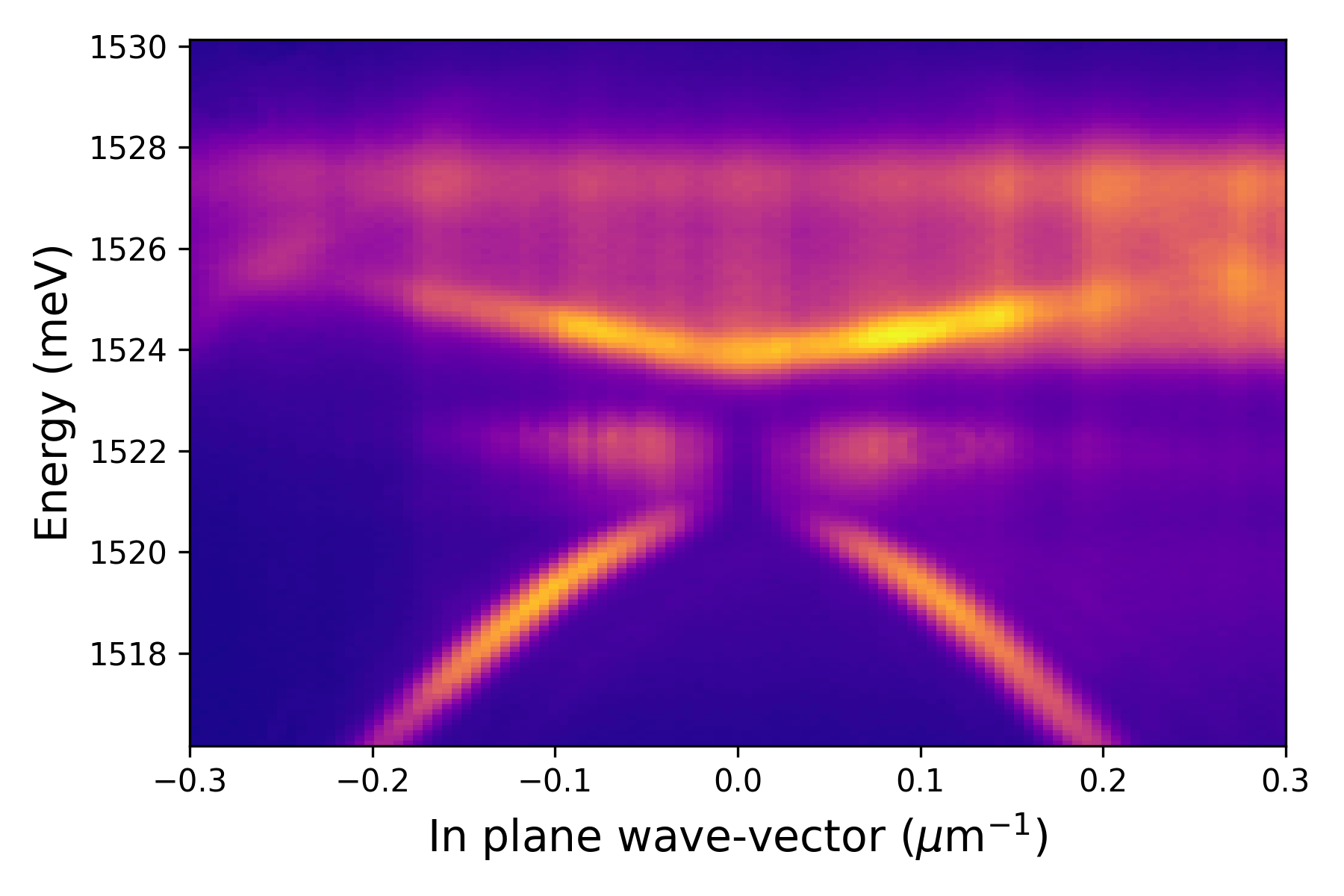}
    \caption{}
\end{subfigure}
\begin{subfigure}[t]{0.3\textwidth}
    \includegraphics[scale = 0.36]{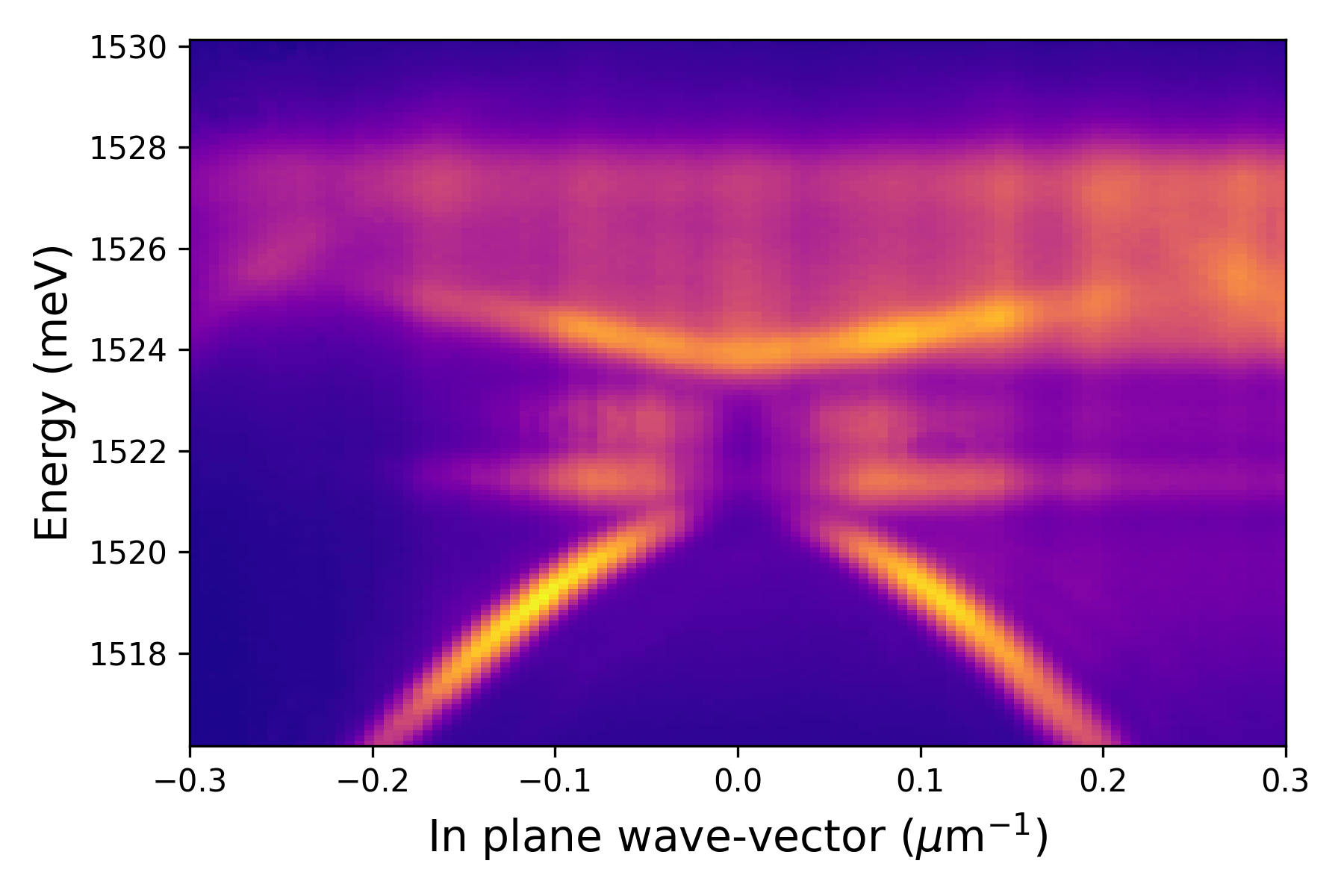}
    \caption{}
\end{subfigure}\\
\vspace{1cm}
\begin{subfigure}[t]{0.3\textwidth}
    \centering
    \includegraphics[scale=0.36]{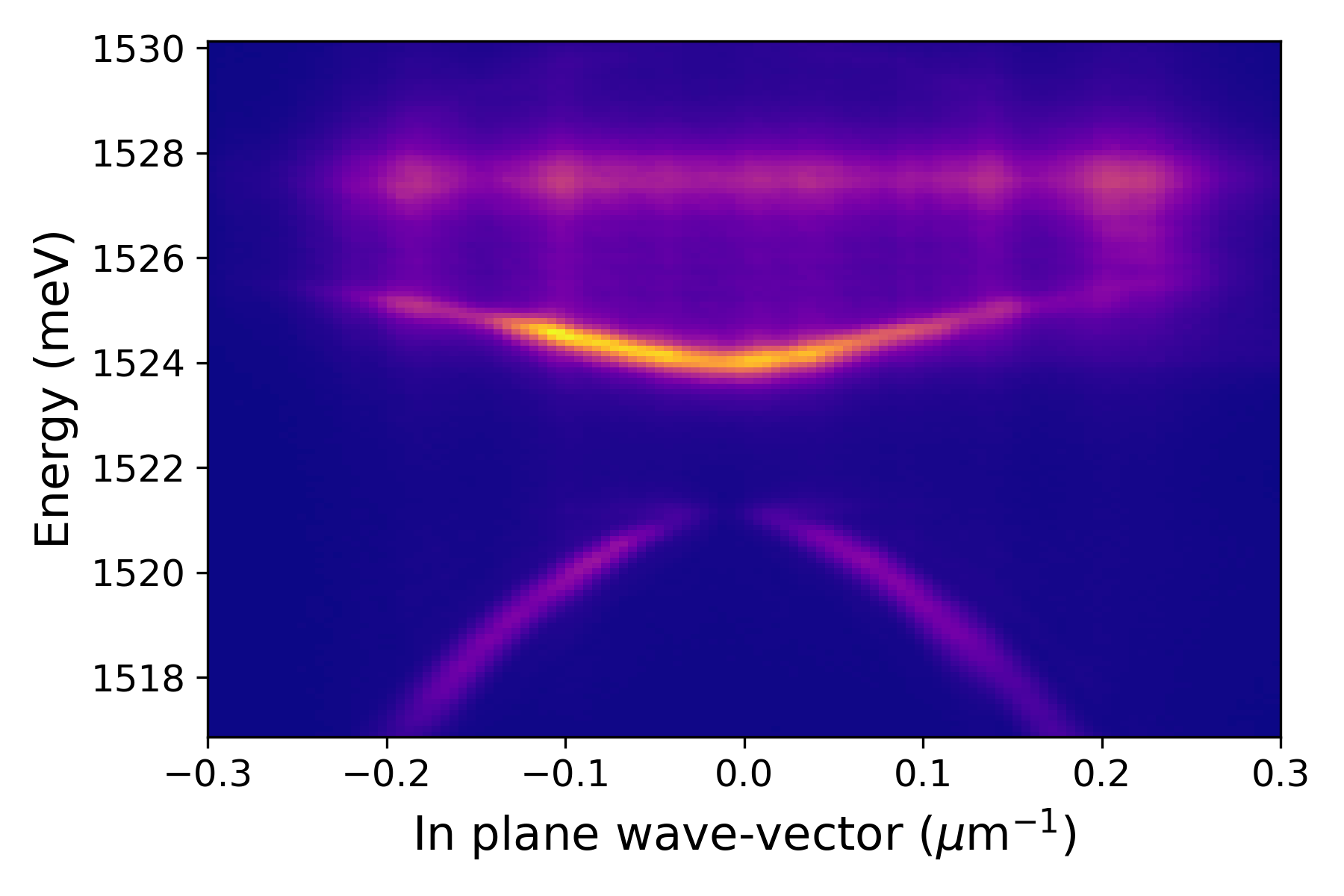}
    \caption{}
\end{subfigure}
\begin{subfigure}[t]{0.3\textwidth}
    \centering
    \includegraphics[scale=0.36]{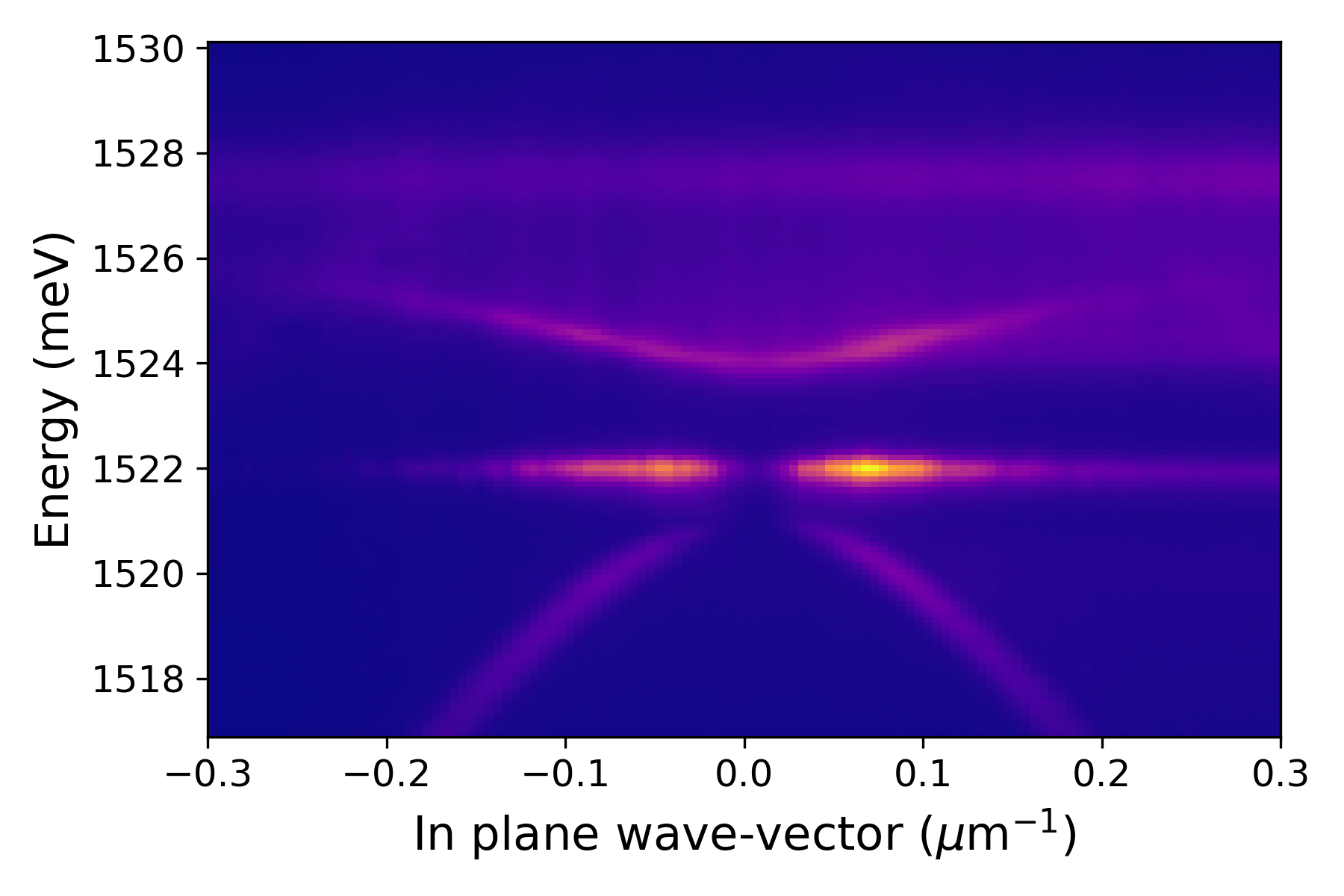}
    \caption{}
\end{subfigure}
\begin{subfigure}[t]{0.3\textwidth}
    \centering
    \includegraphics[scale=0.36]{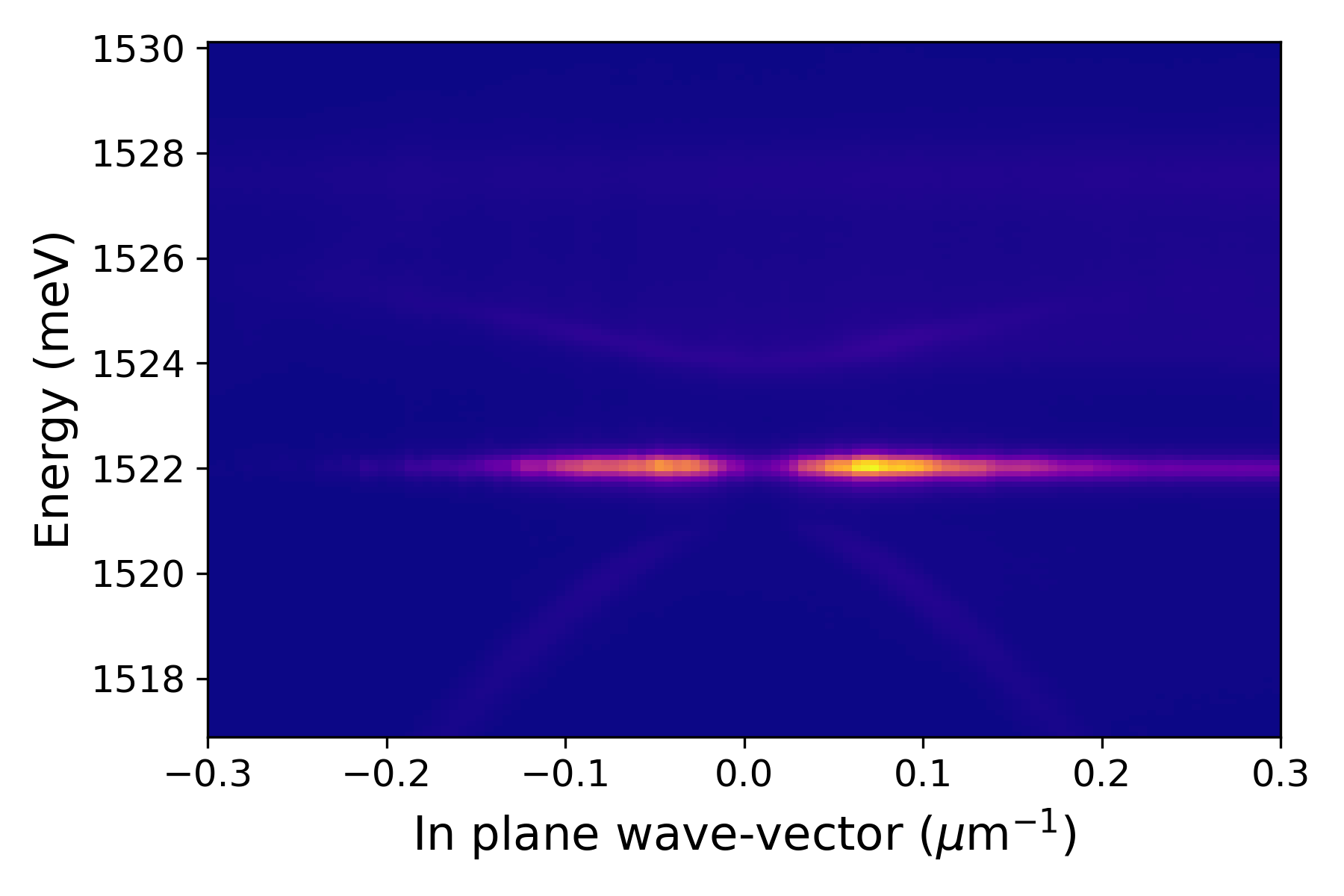}
    \caption{}
\end{subfigure}
\caption{Far field of the polariton emission for different excitation powers for the 130nm deep grating before passivation (a,b,c) and after ALD passivation (d,e,f)}
\label{fig:130nmDisp}
\end{figure}

The ALD processing was proved to be effective in reducing the condensate density, hence the lasing emission from the BIC, by improving the sample's quality. In Fig.\ref{fig:prepostALD_4gratings} we show the emission intensity as a function of the incident laser pumping power before and after aluminum oxide deposition. This was done on different gratings characterized by a pitch = 240nm, depth = 90nm, and filling factors around 70$\%$ - 75$\%$. The results show a reduction in the threshold value of our lasing from BIC that we associate to a lower nonradiative recombination rate of the excitons at the etched sidewalls. 
It is worth noting that in the case of 130nm deep grooves no lasing threshold was observed prior passivation, and the processing was able to restore the sample quality to achieve lasing from a high filling factor structure. Even though not all the structures recovered enough quality to achieve lasing, this result highlights the importance of this post-processing step as displayed in Fig.\ref{fig:130nmDisp}. \\
Despite being a long lifetime state, the lifetime of the photonic bound state in the continuum coupled to a matter state is reduced by the intrinsic excitonic losses, giving rise to the so-called quasi-BIC. As a matter of fact, from theoretical calculations\cite{Ardizzone2021PolaritonContinuum} its lifetime only depends on the BIC excitonic fraction and nonradiative losses. For this reason the state acquires an intrinsic finite lifetime, which can be increased by post-processing the sample. In this sense, the presence of the aluminum oxide layer reduces the density of defects where excitons can decay nonradiatively at the grooves. This step allows to reduce the lasing threshold and also to increase the exciton-polariton lifetime, as demonstrated in the main text. Even though the lifetime of the quasi-BIC is finite, its linewidth always appears to be zero since it cannot couple to the far field. 
\newpage
\section{Cathodoluminescence}
\begin{figure}
\centering
\begin{subfigure}[t]{0.452\textwidth}
    \includegraphics[scale = 0.148]{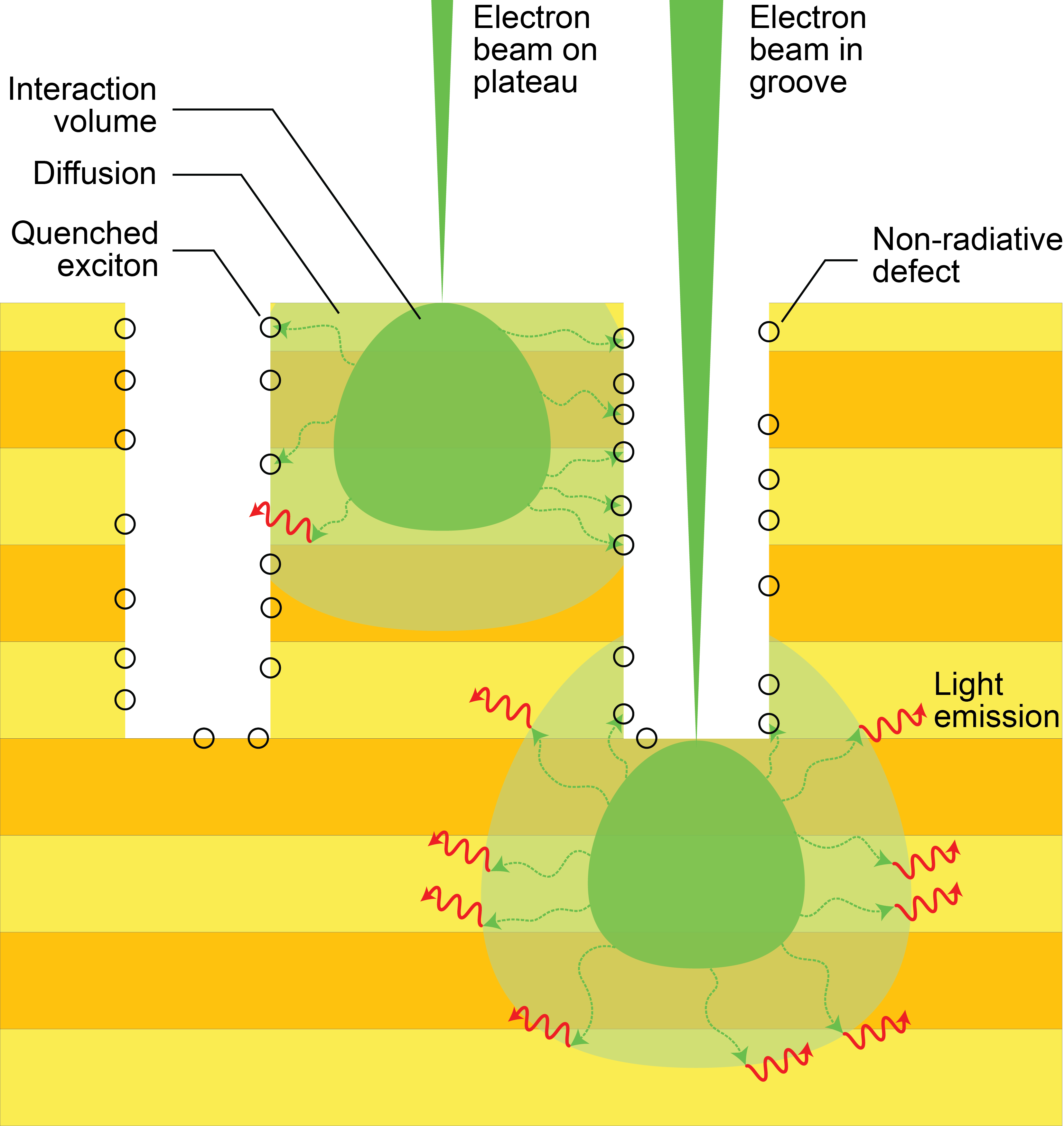}
    \caption{}
\end{subfigure}
\begin{subfigure}[t]{0.54\textwidth}
    \centering
    \includegraphics[scale=0.42]{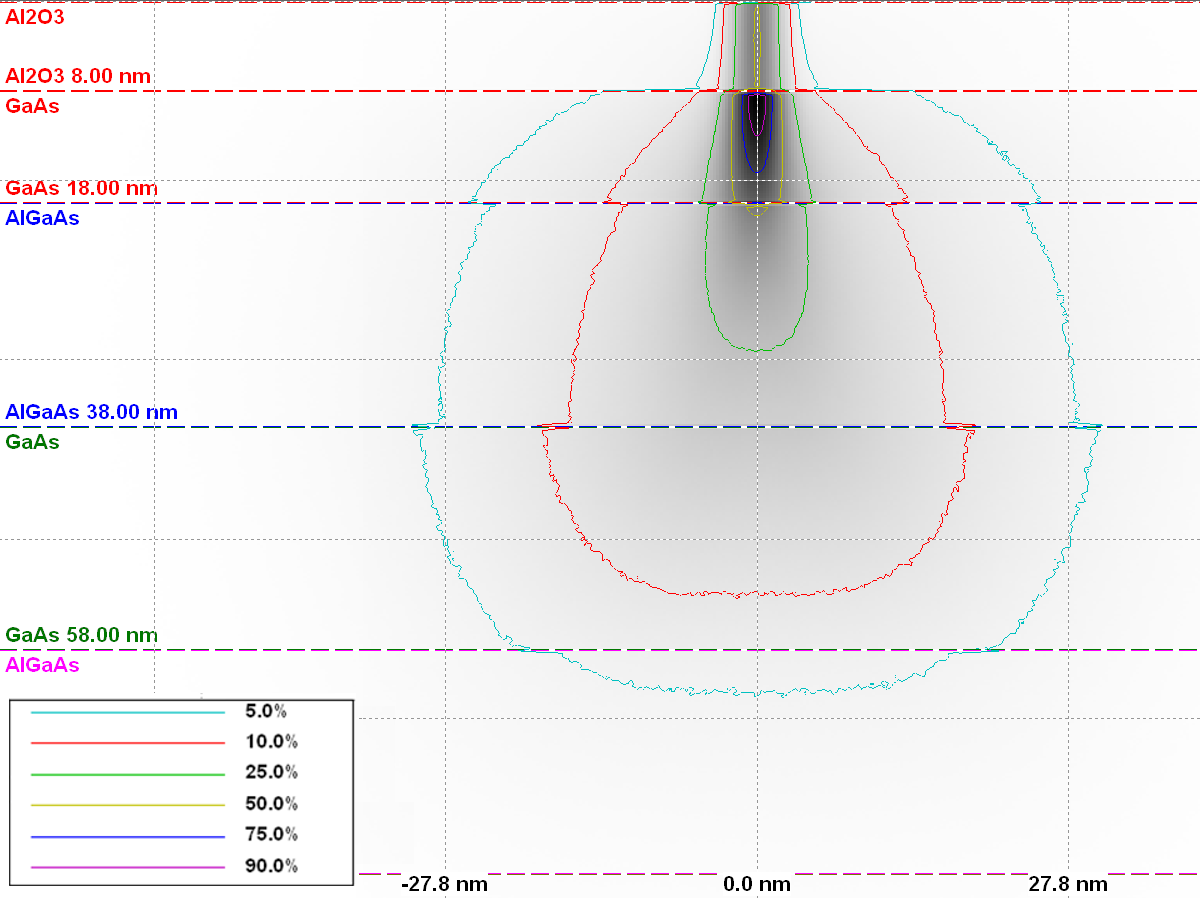}
    \caption{}
\end{subfigure}
\caption{(a) Sketch of the cathodoluminescence interaction with the heterostructure. A low-voltage (3kV) electron beam interacts within a small volume of the sample (dark green) generating excitons and free carriers that diffuse (lighter green). The light emitted is collected by a parabolic mirror and sent to a spectrometer. Part of the generated carriers recombine at surface defects induced by the etching. The two beams shown here represent two possible different excitation positions that occur while the beam scans the sample surface: the beam incident on the plateau probes a larger portion of the surface defects compared to a beam in the groove. As a consequence, a larger part of carriers recombine radiatively in the second case. (b) Monte Carlo simulation of the deposited energy distribution for the 3kV electron beam penetrating the heterostructure. }
\label{fig:quenching}
\end{figure}

Hyperspectral cathodoluminescence (CL) microscopy at 15K was used to investigate the role of etching depth on the excitonic losses. This technique uses electrons from a scanning electron microscope (SEM) to locally excite the sample, which is kept at 15K during the measurements. The emission is collected by a parabolic mirror, which sends the exciton light towards a spectrometer, allowing the reconstruction of a hyperspectral image. As electrons pass through the layers of the heterostructure, they undergo inelastic scattering events that excite a broadband of energy states. Carriers then relax and contribute to the emission from the QWs. The energy of the primary beam was tuned to 3keV to ensure high spatial resolution and surface sensitivity. 
The electron acceleration energy controls the size of the interaction volume of primary electrons in the sample: the lower the energy, the smaller the interaction volume. The interaction volume of a 3kV electron beam reaches depths of few tens of nm from the top surface of GaAs. The electron beam penetration and interaction with the heterostructure was simulated using CASINO Monte Carlo simulations\cite{Hovington1997CASINO:Program,Drouin2007CASINOUsers} (Fig.\ref{fig:quenching}b). At the same time, an electron beam probing a small volume also provides a better lateral spatial resolution. We used this technique to measure the emission from the 40nm and 130nm deep gratings after ALD processing. In both cases, only the first QW is populated directly through electron interaction, while the second one can be populated through carriers photogenerated in the AlGaAs barrier. Other QWs can be indirectly populated with the exciton emission from the first two. This process of exciton passage among QWs is the same that is responsible for the achievement of strong coupling under optical pumping. When excitons are created between the grooves, they diffuse through the QWs to the sidewalls where they may decay in a non-radiative way on etching-induced surface states\cite{Glembocki1995EffectsSurface,Leonhardt1998SurfaceArsenide,Eddy1997GalliumProcess}. Because some of the excitations can effectively change the QW in which they are located, the non-radiative recombination rate will be more important if the etching is deeper.

\begin{figure}
\centering
\begin{subfigure}[t]{0.32\textwidth}
    \centering
    \includegraphics[scale=0.6]{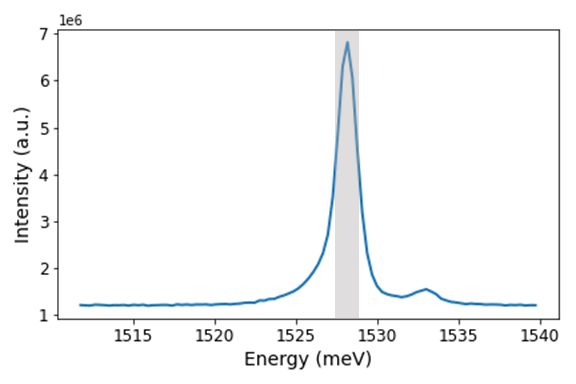}
    \caption{}
\end{subfigure}
    \hspace{1mm}
\centering
\begin{subfigure}[t]{0.31\textwidth}
    \centering
    \includegraphics[scale=0.38]{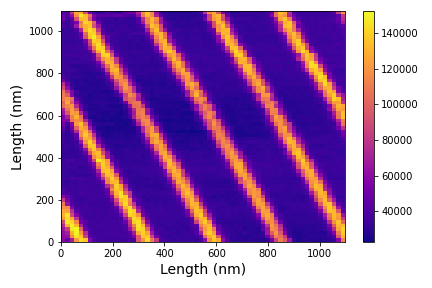}
    \caption{}
\end{subfigure}
    \hspace{1mm}
\begin{subfigure}[t]{0.31\textwidth}
    \centering
    \includegraphics[scale=0.38]{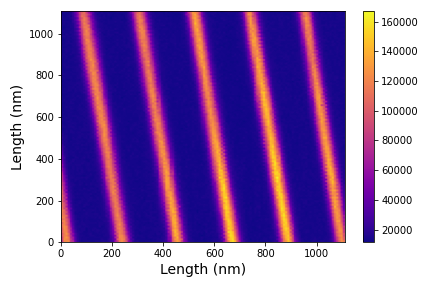}
    \caption{}
\end{subfigure}
\caption{(a) CL emission spectrum under a 3kV electron beam. The gray area shows the energy range considered in the hyperspectral maps (b,c) which corresponds to the exciton energy position. (b,c) Hyperspectral maps of 40nm (b) and 130nm (c) deep grating after ALD processing.}
\label{fig:CL1}
\end{figure}

The surface damage is a limitation of the current waveguide approach which can be addressed via surface treatment. A deeper groove generates a wider surface area through etching, thus increasing the number of non-radiative recombination channels, that is, the system losses in the excitonic component. Therefore, the depth of the grooves can play an important role in the achievement of lasing. In Fig\ref{fig:CL1} we show two hyperspectral maps of the CL emission from a sample with 40nm-deep grooves (b) and 130nm-deep grooves (c).
The intensity of each pixel depends on the probability that excitons excited in that localized position recombine through radiative channels while diffusing away from the interaction volume. If excitons generated at a certain position find non-radiative recombination channels along their random path, the overall intensity associated with that pixel decreases. In both maps in Fig\ref{fig:CL1} we observe that excitation inside the grooves always correspond to a higher CL signal compared to the emission collected when the electron beam hits the plateaus (flat area between two grooves). Finite difference time domain simulations (not shown) discard that this modulation of the CL intensity is due to optical interference effects of etched patterned on the emitted light. We rather hypothesize that the etched sidewalls of the plateaus contain a large number of non-radiative states that quench the exciton emission.  
We observe that the CL intensity modulation depends on the grooves depth. We thus calculated the ratio of CL intensity of the pixels in the grooves over the pixels on the plateaus. The post-processed sample shows that for the 40nm deep grooves, such ratio is around 4.94 (Fig\ref{fig:CL1}.b), meaning that, on average, the light coming from excitation in the grooves is almost five times more intense than the light collected when exciting on the plateaus. For the 130nm case, this ratio reaches values up to 13.22 (Fig\ref{fig:CL1}.c).
We suggest that a deeper etching exposes a larger surface area to the presence of non-radiative defects, thus increasing the probability that diffusing charge carriers recombine at the surface in a non-radiative way. Indeed, the intensity from bottom grooves is comparable to the intensity from unetched areas.

\end{document}

%% file: main.bbl
%